\documentclass[Afour,sageh,times]{sagej}

\usepackage{moreverb,url}

\usepackage[colorlinks,bookmarksopen,bookmarksnumbered,citecolor=red,urlcolor=red]{hyperref}

\newcommand\BibTeX{{\rmfamily B\kern-.05em \textsc{i\kern-.025em b}\kern-.08em
T\kern-.1667em\lower.7ex\hbox{E}\kern-.125emX}}

\def\volumeyear{2016}
\usepackage{graphicx}
\usepackage{subcaption}
\usepackage{algorithm}
\usepackage[noend]{algpseudocode}
\usepackage[utf8]{inputenc}

\newcommand{\lbmpy}{{\em lbmpy}}
\newcommand{\pystencils}{{\em pystencils}}
\newcommand{\walberla}{\textsc{waLBerla}}

\newcommand{\nvidia}{NVIDIA}
\newcommand{\amd}{AMD}

\usepackage{xcolor}
\definecolor{lssblue}{rgb}{0.121,0.231,0.4}
\definecolor{lssred}{rgb}{0.76,0.3,0.19}

\begin{document}

\runninghead{Philipp Suffa, Markus Holzer et al.}

\title{Architecture Specific Generation of Large Scale Lattice Boltzmann Methods for Sparse Complex Geometries}

\author{Philipp Suffa\affilnum{1}, Markus Holzer\affilnum{2}, Harald Köstler\affilnum{1,3} and Ulrich Rüde\affilnum{1,2}}

\affiliation{
\affilnum{1}Chair for System Simulation, Friedrich Alexander Universität Erlangen-Nürnberg, Erlangen, Germany\\
\affilnum{2}Parallel Algorithm Team, CERFACS, Toulouse, France\\
\affilnum{3}Erlangen National High Performance Computing Center (NHR@FAU), Erlangen, Germany\\
}

\corrauth{}

\email{philipp.suffa@fau.de}

\keywords{Sparse Lattice Boltzmann Method, Indirect Addressing, Complex Geometries, High Performance Computing, GPU computing, Large Scale, Porous Media}

\begin{abstract}
We implement and analyse a sparse / indirect-addressing data structure for the Lattice Boltzmann Method to support efficient compute kernels for fluid dynamics problems with a high number of non-fluid nodes in the domain, such as in porous media flows.
The data structure is integrated into a code generation pipeline to enable sparse Lattice Boltzmann Methods with a variety of stencils and collision operators and to generate efficient code for kernels for CPU as well as for \amd{} and \nvidia{} accelerator cards. 
We optimize these sparse kernels with an in-place streaming pattern to save memory accesses and memory consumption and we implement a communication hiding technique to prove scalability.
We present single GPU performance results with up to 99\% of maximal bandwidth utilization. 
We integrate the optimized generated kernels in the high performance framework \walberla{} and achieve a scaling efficiency of at least 82\% on up to 1024 \nvidia{} A100 GPUs and up to 4096 \amd{} MI250X GPUs on modern HPC systems.
Further, we set up three different applications to test the sparse data structure for realistic demonstrator problems. We show performance results for flow through porous media, free flow over a particle bed, and blood flow in a coronary artery. 
We achieve a maximal performance speed-up of 2 and a significantly reduced memory consumption by up to 75\% with the sparse / indirect-addressing data structure compared to the direct-addressing data structure for these applications. 
\end{abstract}

\maketitle

\section{Introduction}
The increasing power of high-performance computing (HPC) systems enables computational fluid dynamic (CFD) simulations, which were still out of scope some years ago. 
The leading HPC systems in the Top500 list\endnote{Top500 List: \url{https://www.top500.org/}, accessed: 2024-6-3} reach a peak performance of ExaFLOPs, so they can perform $10^{18}$ floating point operations per second. This trend is also caused by the utilization of accelerators, such as NVIDIA, AMD, or INTEL Graphics Processing Units (GPUs). 
With these new computing capabilities, computational fluid dynamic (CFD) problems, which were out of scope before, can now be tackled, such as fully resolved porous media simulations at relevant scales, as presented in \cite{mattilaProspectComputingPorous2016a} or entire body arterial flows as in \cite{randlesMassivelyParallelModels2015}. 
However, it is not trivial to fully utilize the maximum performance of these HPC systems, especially for systems with accelerators (\cite{brodtkorbGraphicsProcessingUnit2013}, \cite{hijmaOptimizationTechniquesGPU2023}, \cite{laiHybridMPICUDA2020}, \cite{rakParallelProgrammingHybrid2024}). 

The Lattice Boltzmann method (LBM) (\cite{chenLATTICEBOLTZMANNMETHOD1998a}, \cite{lbm_book}) is an efficient inherently parallel method to solve CFD problems for complex geometries. 
On many HPC systems, the LBM shows excellent performance as presented in \cite{liuAcceleratingLargeScaleCFD2023}, \cite{spinelliHPCPerformanceStudy2023}, \cite{watanabePerformanceEvaluationLattice2022} or \cite{godenschwagerFrameworkHybridParallel2013}.

There are two common ways to store data for LBM simulations. The direct-addressing LBM stores and computes all cells of the domain. We call this technique "dense" or "direct-addressing" data structure in the following. 
It is shown to be very fast and efficient for most simulation setups (see \cite{kummerlanderImplicitPropagationDirectly2023}, \cite{lehmannAccuracyPerformanceLattice2022} and \cite{lattPalabosParallelLattice2021}). 
However, it struggles in performance and memory consumption for simulation domains with a high number of non-fluid nodes, such as in porous media flows. 
In the following, we call such domains "sparse domains". 
In contrast, simulations with a high percentage of fluid nodes, such as a free channel flow, we will refer to as "dense domain".

The second way to store data for LBM simulations is the indirect-addressing storage format. 
We call it "sparse" data structure in the following.
This approach stores only fluid cells and can therefore save a significant amount of memory. Additionally, the sparse approach reaches superior performance for sparse complex geometries as compared with the dense approach. 
In this sense, the sparse approach has been found to be very efficient, see e.g. in \cite{wittmannComparisonDifferentPropagation2013} and \cite{schulzParallelizationStrategiesEfficiency2002}. 
In particular, the sparse data structure is employed successfully to simulate porous media flows as in \cite{panHighperformanceLatticeBoltzmann2004}, \cite{zeiserBENCHMARKANALYSISAPPLICATION2009}, \cite{wangDomaindecompositionMethodParallel2005} and \cite{vidalImprovingPerformanceLarge2010}.

In this work we study the code generation for highly efficient LBMs and their performance on large HPC systems.
The sparse data structure is realised with the code generation framework of \lbmpy{} (\cite{lbmpy}), allowing it to run on a variety of architectures, such as all common CPUs as well as \nvidia{} and \amd{} accelerators. 
The generated sparse compute kernels are integrated in the multiphysics HPC framework \walberla{} (\cite{waLBerla}) to enable massively parallel simulations with excellent scalability (\cite{holzerDevelopmentCentralmomentPhasefield2024}).

We compare the performance of the generated sparse kernels with the dense approach and present the scaling performance of the sparse data structure on modern HPC systems such as JUWELS Booster (\cite{juwels}) and LUMI\endnote{LUMI Supercomputer: \url{https://www.lumi-supercomputer.eu/}, accessed: 2024-3-13}. 
Further, we show performance results for realistic model problems such as a flow in a porous media, a flow over a packed bed, and a coronary artery flow on a high number of accelerator cards.

\section{Lattice Boltzmann Method}
The lattice Boltzmann method is a mesoscopic method established as an alternative to classical Navier-Stokes solvers (\cite{lbm_book}). 
The simulation domain is usually discretized by a lattice of square cells. 
A cell at position $\textbf{x}$ stores a particle distribution function (PDF) $f_i(\textbf{x},t)$, which represents the probability of particles at time $t$ with discrete velocity $\textbf{c}_i$. 
The macroscopic quantities, lattice density $\rho$ and momentum density $\rho \textbf{u}$, can be computed from the PDFs using 
\begin{equation}
    \rho (\textbf{x},t) = \sum_i f_i (\textbf{x},t) \ \ \  \text{and} \ \ \ \rho \textbf{u}(\textbf{x},t) = \sum_i \textbf{c}_i f_i (\textbf{x},t).
\end{equation}
A standard set of discrete three-dimensional velocity directions would be the D3Q19 stencil, which results in $Q = |\{\textbf{c}_i\}| = 19$ PDFs, respectively. 

The Boltzmann equation discretized in time, space, and velocity space reads
\begin{equation}
    f_i(\textbf{x} + \textbf{c}_i \Delta t, t + \Delta t) = f_i(\textbf{x},t) + \Omega_i(\textbf{x},t),
\end{equation}
with $\Delta x$ as lattice spacing, $\Delta t$ as time step size and $\Omega$ as collision operator.
The LB  equation can be separated into a collision step 
\begin{equation}
        \Tilde{f}_i(\textbf{x},t) = f_i(\textbf{x},t) + \Omega_i(\textbf{x},t)
        \label{equ:collide}
\end{equation}
and a streaming step
\begin{equation}
        f_i(\textbf{x} + \textbf{c}_i \Delta t, t + \Delta t) = \Tilde{f}_i(\textbf{x},t),
        \label{equ:stream}
\end{equation}
with $\Tilde{f}_i$ denoting the post-collision state of the PDFs. 

The simplest collision operator is the single relaxation time (SRT) operator 
\begin{equation}
    \Omega_i^{\text{SRT}}(f) = - \frac{f_i - f_i^{\text{eq}}}{\tau} \Delta t,
\end{equation}
which relaxes the PDFs towards the equilibrium $f^{\text{eq}}$ determined by the relaxation time $\tau$.
The equilibrium is given by 
\begin{equation}
    f_i^{\text{eq}}(\textbf{x},t) = w_i \rho \left(1 + \frac{\textbf{u} \cdot \textbf{c}_i}{c_s^2} + \frac{(\textbf{u} \cdot \textbf{c}_i)^2}{2c_s^4} - \frac{\textbf{u} \cdot \textbf{u}}{2c_s^2}\right)
\end{equation}
with the speed of sound $c_s^2 = (1/3)\Delta x^2/\Delta t^2$ and velocity set specific weights $w_i$.
The fluid velocity of a cell at position $\textbf{x}$ is calculated as $\textbf{u}(\textbf{x},t) = \rho \textbf{u}(\textbf{x},t) / \rho (\textbf{x},t)$.
The kinematic viscosity $\nu$ is related to the relaxation time $\tau$ and the dimensionless relaxation parameter $\omega = \frac{\Delta t}{\tau} \in \left]0,2\right[$ by
\begin{equation}
    \nu = c_s^2 (\frac{\tau}{\Delta t} - \frac{1}{2}) = c_s^2 (\frac{1}{\omega} - \frac{1}{2}).
\end{equation}

\section{Data Structures in \walberla{}}
\begin{figure*}[htb]
	\centering
	\includegraphics[width=0.7\linewidth]{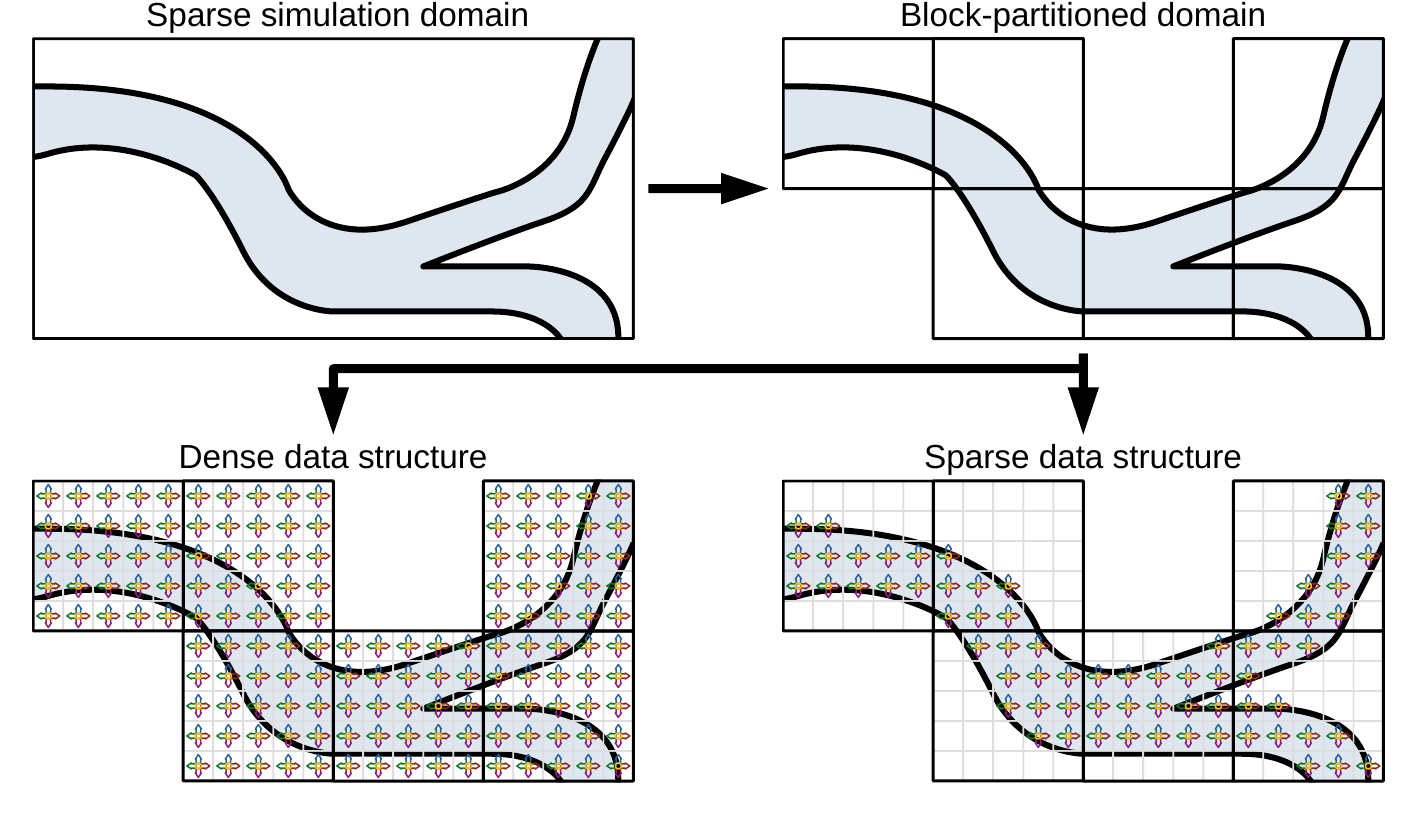}
	\caption{Exemplary setup of a sparse simulation domain in 2D with a low percentage of fluid covering the domain (light blue), and a high number of obstacle cells. Visualisation of the block partitioning with extraction of blocks without fluid. Illustration of a dense and a sparse data structure for an exemplary setup of 5x5 cells per block and a D2Q5 stencil. While the dense data structure stores PDFs and operates on all cells, the sparse data structure only stores and operates on fluid cells.}
	\label{fig:block_decomp}
\end{figure*}

In the \walberla{} framework, the simulation domain is partitioned into uniform cubic blocks, typically with a size of around $64^3$ cells on CPUs and about $256^3$ cells on GPUs. In \autoref{fig:block_decomp} the block partitioning into uniform cubic blocks is shown.
For parts of the domain where no fluid is present, blocks that only consist of obstacle cells can be discarded. 
The remaining blocks are then distributed to the available MPI processes, so that every process gets at least one block. However, more blocks per process are also possible and can be useful, for example, when load balancing is necessary, as we will see in the following. 
The organization of a computational grid into blocks introduces a hierarchy that is found essential for efficient processing on the extreme scale since many operations can be better organized in such a hierarchy. 
In particular, performing the mesh partitioning and load balancing in terms of blocks keeps the complexity and overhead of these algorithms small (see \cite{schornbaumMassivelyParallelAlgorithms2016} and \cite{schornbaumExtremeScaleBlockStructuredAdaptive2018}).

As indicated in \autoref{fig:block_decomp}, the dense / direct-addressing data structure implemented in \walberla{} stores the PDFs for every cell in memory, even for non-fluid cells. 
However, for sparse domains or porous media flows, the porosity 
\begin{equation}
    \phi = \frac{N_{F}}{N}
    \label{equ:porosity}
\end{equation}
with $N$ as the total number of cells and $N_F$ as the number of fluid cells, can be arbitrarily small. 
Therefore, much memory may be wasted by storing non-fluid cells on these blocks. 
Furthermore, a branch statement in the LBM kernel is needed to check, if the current cell is a fluid or an obstacle cell. 
Additionally, non-fluid cells can cause unnecessary memory traffic, when cache lines contain fluid and non-fluid cells, and hardware prefetchers may read data from non-fluid cells, which is not used. 
Especially when the domain is sparse, the dense approach creates a significant overhead and can lead to a significant performance loss (\cite{godenschwagerFrameworkHybridParallel2013}).

\subsection{Sparse Data Structure}
To avoid the disadvantages of the dense data structure, we have developed a sparse data structure in \walberla{} and \lbmpy{}. 
This technique is also used by other LBM frameworks such as HARVEY (\cite{randlesPerformanceAnalysisLattice2013a}), Musubi (\cite{hasertComplexFluidSimulations2014}), ILBDC (\cite{zeiserBENCHMARKANALYSISAPPLICATION2009}) and MUPHY (\cite{bernaschiMUPHYParallelHigh2008}), just to mention a few.

The idea is to only store fluid cells in a one-dimensional array, we call \texttt{PDF-list}, so that no memory is wasted on storing non-fluid cells. 
Furthermore, with such a data structure LBM kernels only have to iterate over fluid cells, and no branch conditions are needed in the innermost kernel loops. 
On the other hand, one loses spatial information when storing cell data in a linear array only. 
With the direct-addressing data structure, PDFs can easily be accessed by their spacial location $\textbf{x}$ and the PDF index $i$. 
This access via index arithmetic is not possible for the sparse \texttt{PDF-list}. 

A second data structure, the \texttt{index-list}, is introduced to recover the lost spacial information. 
This list stores the streaming information from one PDF to another. 
So for one PDF, the \texttt{index-list} stores the location of the PDF, to which it will propagate to in the streaming step. 
Using the \texttt{index-list}, we can access the neighbors of a cell using one indirection. 
Therefore, this approach is called an indirect-addressing scheme.

While the \texttt{PDF-list} consists of $N_F \cdot Q$ entries, where $Q$ is the size of the stencil, the \texttt{index-list} only consists of $N_F \cdot (Q-1)$ entries, since the center PDF need not to be stored, as no propagation information is needed for the center PDF.

\begin{figure}[htb]
	\centering
      \begin{subfigure}{0.5\textwidth}
         \centering
         \includegraphics[page=1, width=\linewidth,trim={7.5cm 1.5cm 7.5cm 2cm},clip]{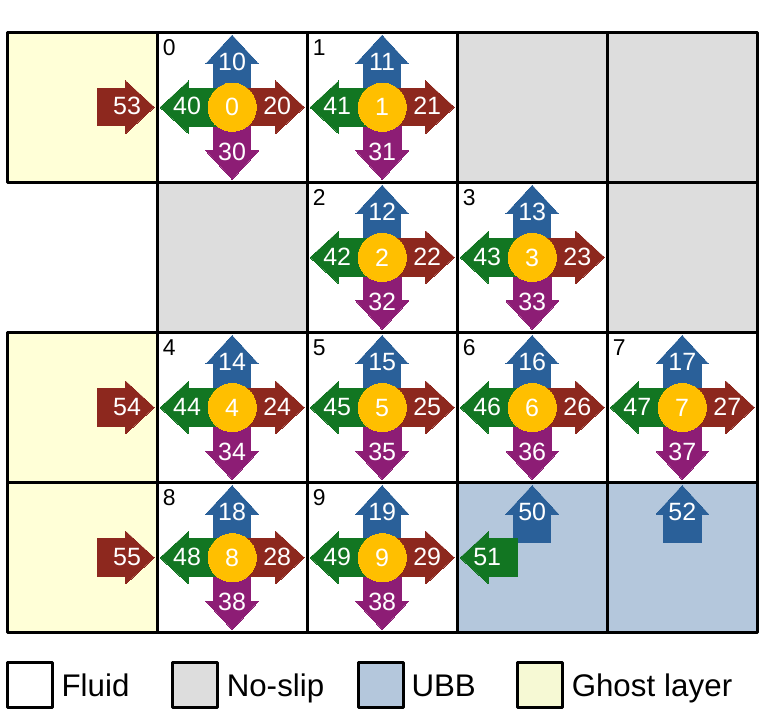}
         \subcaption{\texttt{PDF-list} in SoA layout with appended PDFs from velocity-bounce-back (UBB) boundaries (in light blue) and ghost layers (in light yellow). The numbers indicate the position of the PDF in the \texttt{PDF-list}.}
         \label{fig:sparseStructurePDFList}
     \end{subfigure}
     \begin{subfigure}{0.5\textwidth}
         \centering
         \includegraphics[page=2, width=\linewidth,trim={7.5cm 3cm 7.5cm 2cm},clip]{SparseData.pdf}
         \subcaption{Structure of the \texttt{index-list} for the corresponding \texttt{PDF-list} in \autoref{fig:sparseStructurePDFList}. The numbers indicate from which position in the \texttt{PDF-list} to pull from in the streaming step.}
         \label{fig:sparseStructureIndexList}
     \end{subfigure}
     \begin{subfigure}{0.5\textwidth}
         \centering
         \includegraphics[page=3, width=\linewidth,trim={5.3cm 10cm 6cm 1.5cm},clip]{SparseData.pdf}
         \subcaption{Actual structure of \texttt{PDF-list} and \texttt{index-list}, as they lie in memory in SoA layout.}
         \label{fig:sparseStructureLists}
     \end{subfigure}
    \caption{Structure of the \texttt{PDF-list} and the \texttt{index-list} for an exemplary D2Q5 velocity set. The domain contains fluid cells (white), ghost layers (light yellow), velocity-bounce-back (UBB) boundaries (light blue) and no-slip boundaries (grey). The directions of the PDF stencil are indicated by colors as well. In direction west there is a MPI interface to the neighboring block considered. North, east and south cells next to the presented cells are also considered as no-slip cells.}
	\label{fig:sparseStructure}
\end{figure}

The exact structure of the \texttt{PDF-list} and \texttt{index-list} is illustrated in \autoref{fig:sparseStructurePDFList} and \autoref{fig:sparseStructureIndexList}, respectively. 
We show the \texttt{PDF-list} in a Structure-of-Array (SoA) format, so all PDFs of one direction lie next to each other in memory.
The demonstrator domain consists of fluid cells (white), no-slip boundary cells, which indicate an obstacle (grey), velocity bounce back boundary conditions (blue) and ghost layer cells, which are needed for the communication between blocks (light yellow).
 
In \autoref{fig:sparseStructurePDFList} the PDF of cell 0 in direction west ($PDF_w^0$) is stored at position 40 in the \texttt{PDF-list}, and the $PDF_w^1$ of cell 1 is stored at position 41. 
To perform a streaming step for $PDF_w^0$ in cell 0, we have to look up the pull index (for a presumed pull-streaming pattern) in the \texttt{index-list}. 
This is illustrated in \autoref{fig:sparseStructureIndexList}, where the pull index of the $PDF_w^0$ is the PDF 41. This makes sense, as for direction west we pull $PDF_w$ from the right neighbour cell. 
The pull index look-up in the \texttt{index-list} is done for all PDFs of all fluid cells to perform a complete streaming step.
The actual layout of the \texttt{PDF-list} and \texttt{index-list} in memory is illustrated in \autoref{fig:sparseStructureLists}. There the SoA layout is used.

\subsubsection{Sparse Boundary Conditions}
Some modifications to the list data structures are made to support the implementation of boundary conditions. No-slip boundary conditions can easily be realised by setting the pull indices of the PDF, which would pull from a no-slip boundary to the inverse direction of the PDF. This is also illustrated in \autoref{fig:sparseStructureIndexList}. 
For illustration we focus on $PDF_w^1$ of cell 1. It would pull from its right neighbour cell, but this is an obstacle (no-slip) cell. 
So $PDF_w^1$ (PDF 41) pulls from the PDF in east direction of its same cell 1, which is then the $PDF_e^1$ with index 21. 
Also, periodic boundary conditions are easy to implement; here, the pull index of the PDF, which must stream from the periodic boundary on the opposite side, is just set to the PDF on the other side of the domain.

For boundary conditions other than no-slip or periodic, PDFs, which correspond to a boundary cell but point to a fluid cell, must be appended to the \texttt{PDF-list}. Further, the \texttt{index-list} has to be modified, so that PDFs of fluid cells next to boundary cells pull from the boundary PDF cells. 
This is also illustrated for velocity-bounce-back boundary conditions (UBB) in \autoref{fig:sparseStructure}. 
Here, the $PDF_w^9$ (PDF 49, cell 9, direction west) pulls from $PDF_w$ of the UBB boundary cell right next to it, which is the appended PDF with index 51.

\subsubsection{Sparse Communication}
For the dense data structure, every block has a ghost layer of at least one cell all around in which it stores the PDFs traveling in the corresponding direction. 
This ghost layer is used to communicate PDF information between MPI processes. 
For the communication between sparse blocks, we also have to append these ghost layer PDFs to the \texttt{PDF-list} and modify the \texttt{index-list} so that cells next to the MPI interface pull from these ghost layer PDFs (see \autoref{fig:sparseStructure} yellow cells).

Nevertheless, as we only append boundary and ghost-layer PDFs pointing to fluid, the memory overhead of the additionally stored PDFs is relatively low. In LBM kernels, we still only need to iterate over fluid cells.

\section{Code Generation for Sparse Kernels}
Many variants of the LBM have been developed over the last decades, which vary in complexity, accuracy, and computational cost. 
The code generation framework \lbmpy{} is capable to generate kernels for most of these LBMs. 
It supports a wide range of velocity sets, for example, the D2Q5, D2Q9, D3Q15, D3Q19, D3Q27, and more. 
Further, highly efficient code for different collision operators can be generated.
For example, the classical collision models such as single-relaxation time (SRT), two-relaxation time (TRT), and multi-relaxation time (MRT) operators (\cite{lbm_book}) are available. 
However, more advanced collision models are also supported, such as the central moment operator or the cumulant operator. The cumulant LBM, e.g., provides superior accuracy and stability for high Reynolds number flows (see \cite{geierCumulantLatticeBoltzmann2015}). 
The complexity of the collision models increases from the SRT model to the cumulant model in terms of complexity and the number of moment transfers, so e.g. the transfer from moment space to central moment space or to cumulant space. 
This can increase the number of floating point operations in the collision step significantly.
However, due to optimizations such as common sub-expression elimination (CSE), the number of operations per cell lies between only 200 and 400 FLOPS for a D3Q19 stencil irrespective of the collision model (\cite{hennigAdvancedAutomaticCode2022}). 
Therefore, the performance of the compute kernels remains memory-bound, as the number of memory accesses stays constant for all collision operators. 
Consequently, we can report a similar performance for all collision operators in the following in \autoref{fig:CompPullAAGPU}.

Additionally, \lbmpy{} provides a generic development tool to design new collision schemes. 
The high-level domain-specific language of \lbmpy{} allows the user to formulate, extend, and test various Lattice Boltzmann Methods (\cite{lbmpy}).

To profit from the functionalities of the code generation pipeline \lbmpy{}, we integrated the generation of new sparse LBM kernels, boundary handling kernels, and communication kernels.
All together, we are now able to generate efficient sparse kernels for various velocity sets and collision operators, which can run on all common CPUs and \nvidia{} and \amd{} GPUs.

\begin{figure}[htb]
	\centering
    \includegraphics[width=0.8\linewidth]{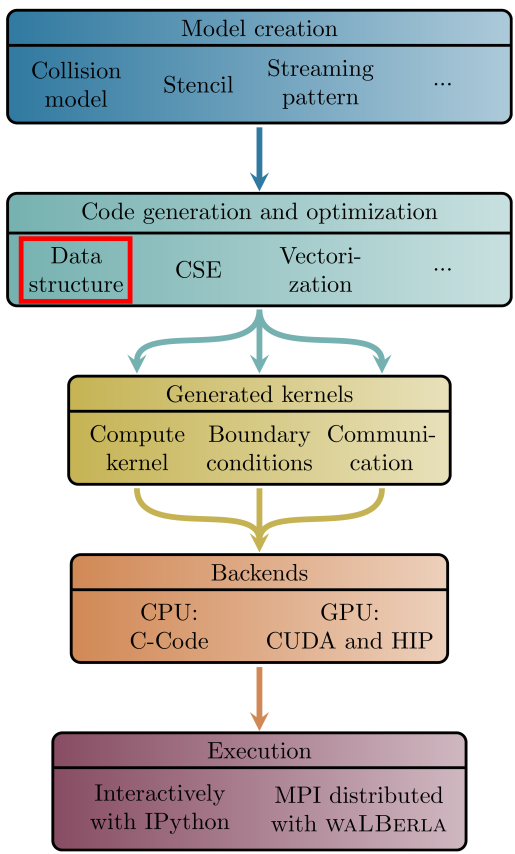}
	\caption{Complete workflow of the code generation pipeline of \lbmpy{}. For a full CFD application compute kernels as well as boundary and communication kernels are generated. }
	\label{fig:lbmpywithwalberla}
\end{figure}

In \autoref{fig:lbmpywithwalberla}, the code generation pipeline is presented. In the model creation, the user defines the LB method by choosing a specific collision model, a stencil and a streaming pattern. Further, a force model, the option of storing the PDFs in a zero-centered storage fashion, or similar can be set. 
After defining the model, the LB method is represented as a set of equations stored in an abstract syntax tree (AST). 
This AST is now passed to \pystencils{} (\cite{Bauer19}), which can perform further optimizations, such as common subexpression elimination, loop splitting, or adding vector intrinsics for single-instruction-multiple-data (SIMD) execution. 
At this point, we have to decide about the data structure of the generated kernels to be sparse or dense, which defines the loop nest of the generated code and the data accesses. 
To support a complete CFD application, the generation of LB kernels, as well as boundary and communication kernels, is needed. The code generation pipeline supports all of the kernels for the sparse and the dense data structure. 
As the last step, the actual code is generated for a specified architecture. 
In particular, all kernels can be generated in plain C-code to support all common CPUs. 
However, the kernels can also be generated with a HIP\endnote{HIP: \url{https://rocm.docs.amd.com/projects/HIP/en/latest/}, accessed: 2024-5-13} or CUDA\endnote{CUDA: \url{https://developer.nvidia.com/cuda-toolkit}, accessed: 2024-5-13 } API to support the GPU venders \amd{} or \nvidia{}, respectively.
Lastly, these optimized kernels can be run interactively in IPython\endnote{IPython: \url{https://ipython.org/}, accessed: 2024-4-15}, but they can also be integrated into a high-performance framework such as \walberla{} to be run in parallel on thousands of CPU or GPU nodes.

\subsubsection{Single Node Results}
\begin{figure}[htb]
	\centering
     \includegraphics[width=1.0\linewidth]{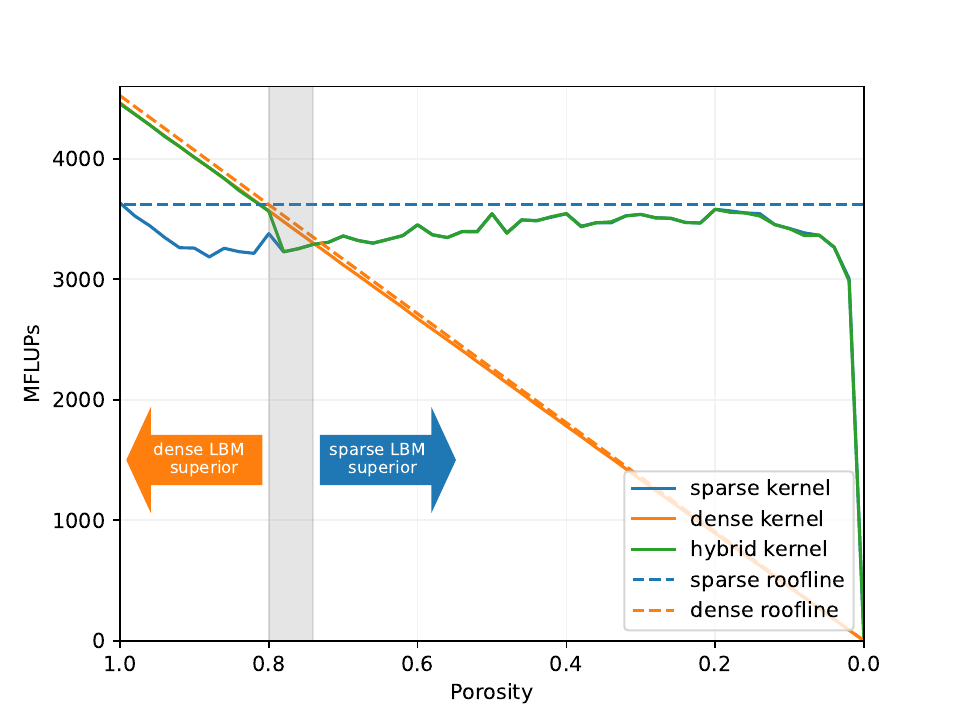}
    \caption{Single GPU benchmark for sparse, dense and hybrid data structure with varying porosity on a \nvidia{} A100 with $256^3$ cells, D3Q19 velocity set and SRT collision operator. The theoretical performance is calculated from the bandwidth of a streaming benchmark (1361 GB/s) and the theoretical number of memory accesses of the kernels, as LBM code is usually memory bound.}
    \label{fig:SparseVsDense}
\end{figure}

In \autoref{fig:SparseVsDense}, we present the performance of the generated sparse LBM kernel in comparison to the generated direct-addressing kernel. 
The diagram shows the mega fluid lattice updates per second (MFLUPs) depending on the porosity $\phi$ as defined in \autoref{equ:porosity}.
The LB method uses a D3Q19 velocity set and the SRT collision operator. The benchmark measures the LBM kernel performance without boundary handling or communication, and it is performed on a single \nvidia{} A100 GPU. 

We observe that the single GPU performance of the sparse kernel is quite close to the theoretical peak performance. 
Furthermore, the MFLUPs performance remains essentially constant for decreasing porosity. 

However, \autoref{fig:SparseVsDense} also shows that the sparse kernels perform worse for $\phi \geq 0.75$. 
This is caused by the extra memory accesses of the sparse data structure. 
A dense kernel has to read and write every PDF of a cell per time step. 
This results in a memory access volume of $2 Q \cdot B_{\text{PDF}}$ bytes per cell on GPUs, with $Q$ as stencil size, here 19, and $B_{\text{PDF}}$ as the bytes per stored PDF, here 8 bytes for double precision. 
The sparse kernel, on the other hand, accesses $2 Q \cdot B_{\text{PDF}} + (Q - 1) \cdot B_{\text{idx}}$ bytes per cell, because it needs to read neighboring information from the \texttt{index-list}. 
$B_{\text{idx}}$ is the number of bytes per index in the \texttt{index-list}, here 4 bytes for an integer.

Nevertheless, the performance for the dense kernel decreases linearly when porosity decreases. 
The reason for this behavior is that dense kernels in \walberla{} traverse all cells, including the non-fluid cells. This avoids the need for a branch instruction for non-fluid cells, as mentioned before. On the other side, this leads to a linear decrease of the fluid lattice updates per second.

As indicated in \autoref{fig:SparseVsDense}, the theoretical break-even point for sparse and dense kernels is approximately $\phi \sim 0.75$. 

\begin{figure}[htb]
	\centering
     \includegraphics[width=1.0\linewidth]{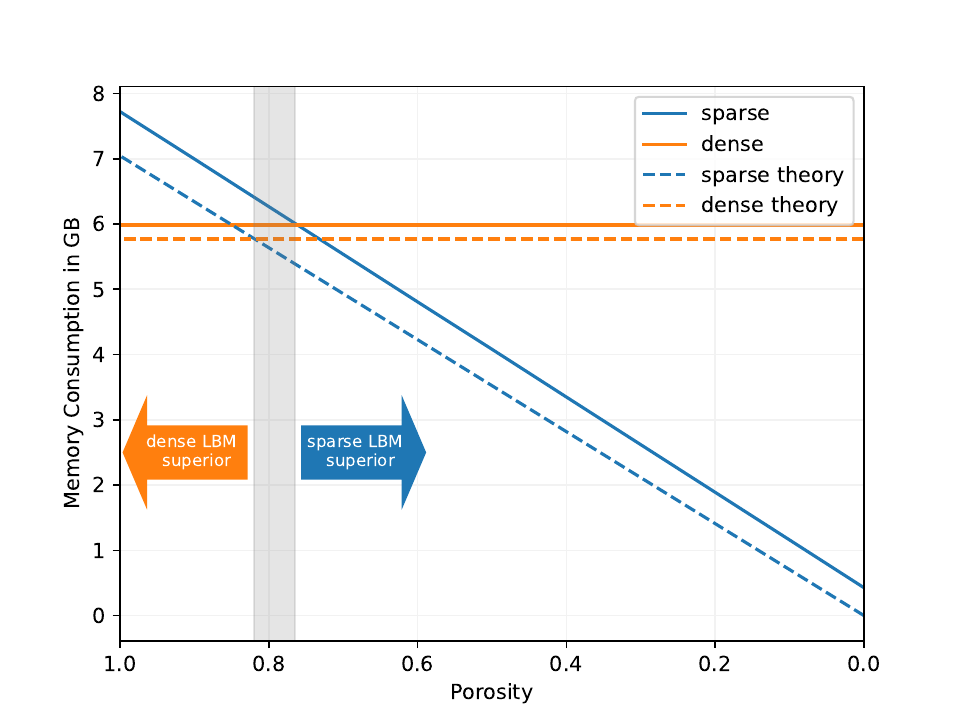}
    \caption{Memory consumption benchmark for $256^3$ cells on a single \nvidia{} A100 GPU for D3Q19 stencil and pull streaming pattern. The theoretical memory consumption is calculated in \autoref{equ:mem}.}
    \label{fig:SparseVsDenseMemory}
\end{figure}

On the same \nvidia{} A100 GPU we present a comparison of the memory consumption for a lattice of $256^3$ cells in \autoref{fig:SparseVsDenseMemory}. 
The memory usage is measured with the \nvidia{} monitoring tool \textit{nvidia-smi}, showing that the sparse data structure consumes linearly less memory with decreasing porosity. 
On the other hand, the dense data structure exhibits a constant memory footprint because it stores all cells in the domain, regardless of whether a cell is fluid or boundary. 
For a porosity of 1.0, where all cells in the domain are fluid, the sparse data structure consumes more memory, since it also has to store the \texttt{index-list} in addition to the \texttt{PDF-list}. 
The theoretical memory consumption of the LBM kernels can be calculated as: 

\begin{align}
    M_{\text{sparse}} &= N_{\text{cells}} \cdot (\underbrace{2 \cdot Q \cdot B_{\text{PDF}}}_\text{PDF-lists} + \underbrace{(Q-1) \cdot B_{\text{idx}}}_\text{index-list} + \underbrace{5 \cdot B_{\text{PDF}}}_\text{other\ fields}) \cdot \phi, \nonumber \\
    M_{\text{dense}} &= N_{\text{cells}} \cdot (\underbrace{2 \cdot Q \cdot B_{\text{PDF}}}_\text{PDF-field} + \underbrace{5 \cdot B_{\text{PDF}}}_\text{other\ fields}).
    \label{equ:mem}
\end{align}

The additional fields stored are a velocity field (3D), a density field (1D), and a flag field to indicate boundary cells (1D). 

We see that for the theoretical as well as for the measured memory footprint, the break-even point of the sparse and dense data structure is at a porosity of around $\phi \sim 0.8$, which is a similar result as for the performance comparison. For a higher porosity, the dense data structure is more suitable in terms of memory consumption, and for a lower porosity, the sparse structure becomes superior.

The measured memory consumption for the sparse as well as for the dense LBM in \autoref{fig:SparseVsDenseMemory} is close to the theoretical memory consumption, so that there is only a small overhead of less than $10\%$ coming from other data structures than the necessary pure PDF data.

\subsection{Hybrid Data Structure}
\label{sec:HybridData}
In certain application scenarios, a hybrid data structure may be advantageous. As an example, consider a free flow over a particle bed as depicted in \autoref{fig:ParticleBedFreeFlow}. After the domain partitioning, some blocks contain only fluid cells, while other blocks consist primarily of non-fluid cells. In this case, neither the sparse nor the dense data structure seems to fit the given scenario perfectly. 

Therefore, we implement the hybrid simulations in \walberla{}. 
From \lbmpy{}, sparse and dense LBM and boundary kernels are generated. 
In \walberla{}, the porosity $\phi$ is calculated individually on each block to determine the block as a sparse or dense block, based on a porosity threshold $\phi_S$. 
Based on the results in \autoref{fig:SparseVsDense} and \autoref{fig:SparseVsDenseMemory}, the porosity threshold should be around $\phi_s \sim 0.8$. 
During the creation of the data structures on the blocks,  a dense PDF field or a sparse \texttt{PDF-list} and the corresponding \texttt{index-list} is created on the block, and only the corresponding generated sparse or dense kernel runs on the blocks.
Besides this functionality, appropriate routines for the communication between sparse and dense blocks must be realized. 
Again, suitable pack and unpack kernels are generated for CPU or GPU architectures with \lbmpy{}, while the MPI communication routine itself stays unchanged. 

In \autoref{fig:SparseVsDense}, we display the performance for the hybrid data structure in green with $\phi_S = 0.8$. As expected, the performance reflects that of the dense kernel for $\phi \geq \phi_S$ and, that of the sparse kernel for $\phi < \phi_S$. 
Consequently, the hybrid approach can always reach the maximum possible \walberla{} performance per block, independent of the porosity. 
The same holds for the memory consumption. If we set the porosity threshold to $\sim 0.8$ as suggested in \autoref{fig:SparseVsDenseMemory}, we also get the best possible memory consumption per block by utilizing the hybrid data structure.

\section{Optimizations to sparse LBM}
The high-performance framework \walberla{} in combination with \lbmpy{} already provides a wide range of optimizations for LB methods. For the sparse data structure, some of the optimizations had to be adapted or re-implemented. In the following, we specifically describe the implementation of an in-place streaming pattern and a communication hiding technique specially designed for the sparse data structures.

\subsection{In-place Streaming: AA Pattern}
\label{sec:AApattern}
\begin{figure*}[htb]
	\centering
    \includegraphics[width=0.8\linewidth]{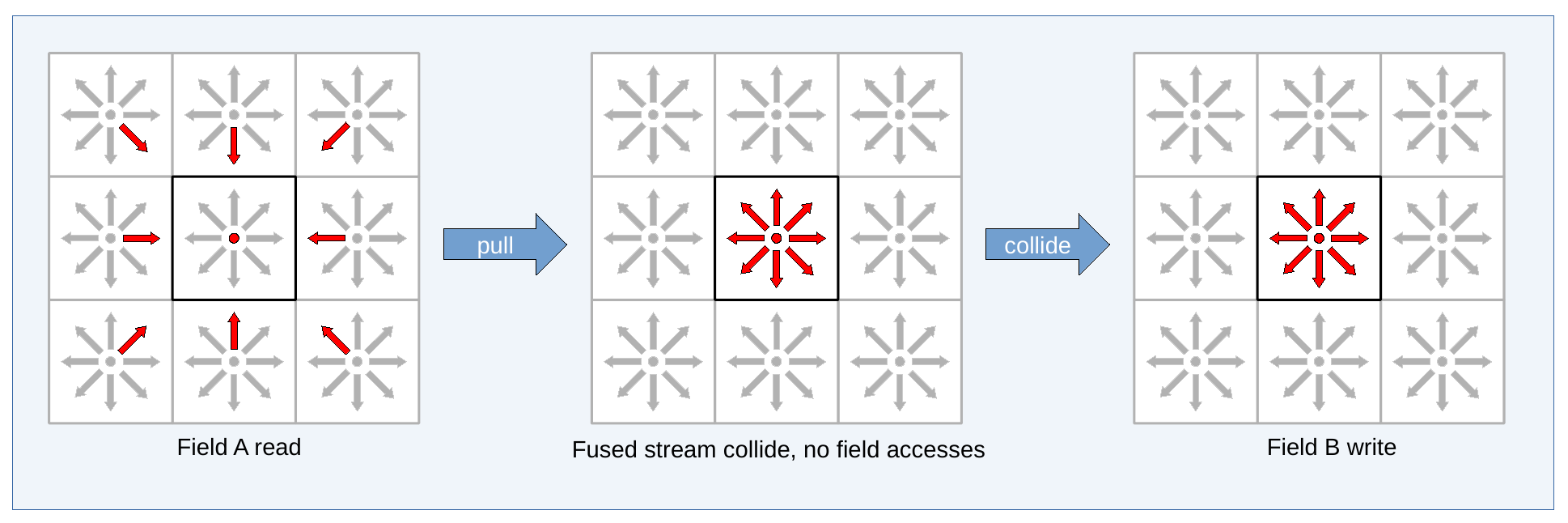}
    \caption{Pull pattern: PDFs are read from field A and, after the fused stream-collide step, they are written to field B.}
	\label{fig:PullScheme}
\end{figure*}

The most common streaming patterns for LBM are the two-grid algorithms, where either PDFs of a cell are pushed into the neighbor cells (push scheme), or PDFs are pulled from the neighbor cells (pull scheme) (\cite{lbm_book}). These algorithms have in common that a temporary PDF field is needed. 
This is because PDFs are stored in a different position than where they are read from. 
As illustrated in \autoref{fig:PullScheme}, these two-grid streaming patterns read from PDF field A, then they propagate (push/pull) the PDFs, and, lastly, they store the propagated results at a different location in the temporary PDF field B. Therefore, these streaming methods are also called AB patterns. After the propagation, a field swap of fields A and B is needed. 

\begin{figure*}[htb]
	\centering
    \includegraphics[width=0.8\linewidth]{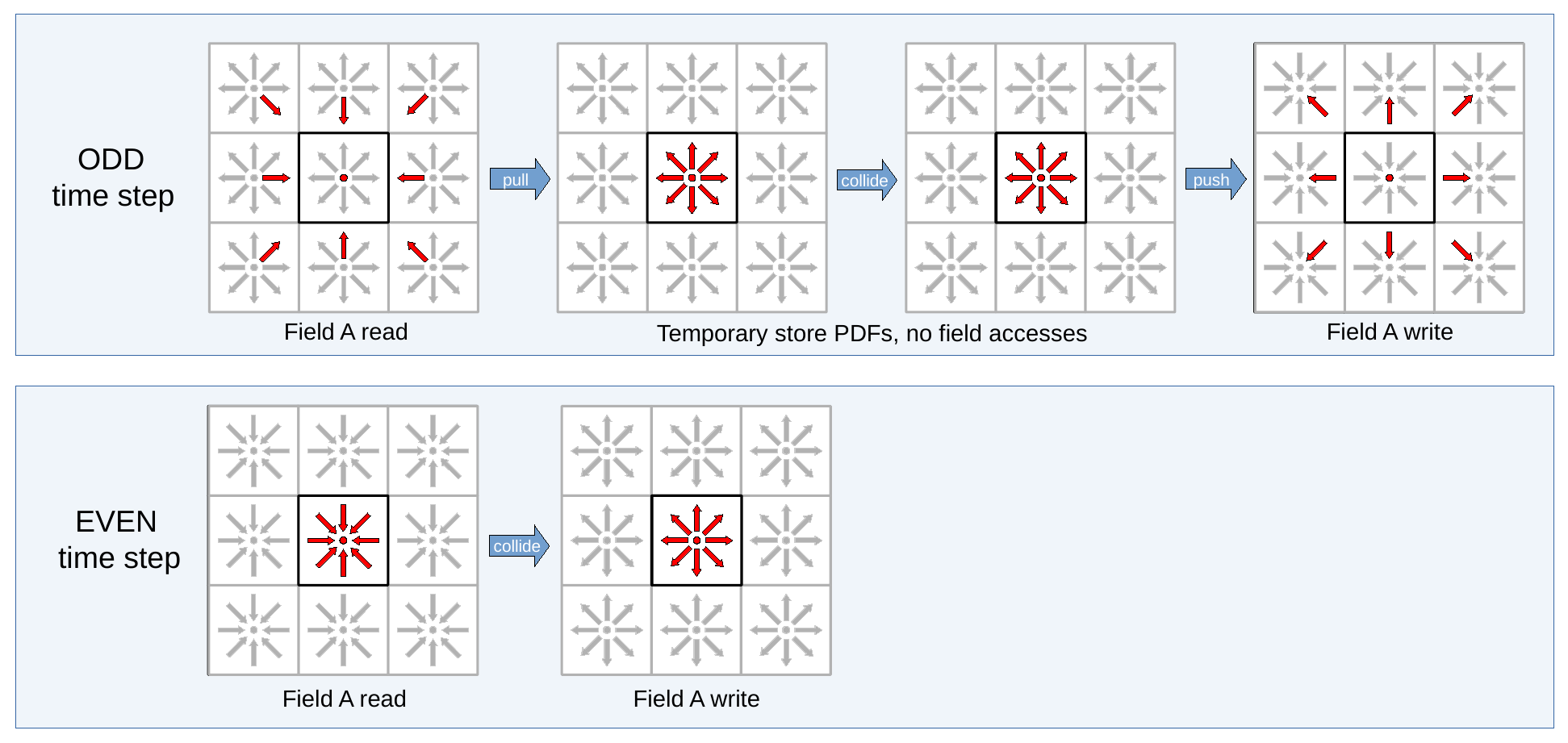}
    \caption{AA pattern: In the "odd" time step, a pull streaming is done, followed by a collision and a push streaming step. As the PDFs are pushed to the same positions as they are read from, the PDFs are stored in the opposite stencil directions. In the "even" time step, only one collision is done, where the PDFs have to be written from the opposite stencil direction to achieve correct macroscopic values. Read and write takes place on the same field, therefore only one PDF field must be stored.}
	\label{fig:AAScheme}
\end{figure*}

The in-place streaming AA pattern on the other hand, enables writing and storing PDF values in the same positions of the PDF field, so that PDFs can be read from field A and also be written to field A without creating data dependencies. 
This saves the memory of the temporary PDF field B. Additionally, it also saves memory accesses (see \cite{baileyAcceleratingLatticeBoltzmann2009}).

These savings are achieved by introducing two alternating streaming time steps, as shown in \autoref{fig:AAScheme}. 
In the "odd" time step, PDFs are read from field A and pulled to the current cell. Then, the collision is performed, and the resulting PDFs are pushed back to the neighbor cells to the positions where they were read from on field A. 
It is worth mentioning, that while we introduced the LBM streaming step as a separate algorithm step in \autoref{equ:stream}, in the actual generated LB kernels streaming and collision is a fused step to save memory accesses.
The odd time step consists of two propagations and one collision, and the reads and writes of the PDFs take place at the same positions on the same field A. 
However, this means that after the second propagation, the PDFs are stored in the wrong position of the stencil. 
Thus we have to take care that PDFs are always in the opposite stencil position after an odd time step. 
It becomes relevant when we must calculate macroscopic values or when we must communicate with neighbor MPI blocks after an odd time step.

The second time step, the "even" time step, only consists of one collision. The PDFs needed for the collision must be read from the opposite stencil positions to get correct macroscopic value calculations and similar. As there is no propagation in the even time step, no neighboring information is needed there. Consequently, there is also no need to access the pull index list, which saves memory access, as discussed in the following. 

After one odd and one even time step, the PDFs are again stored in their right positions, and the outcome is the same as after two push or pull time steps.

The benefit of the AA streaming pattern in terms of memory accesses is shown in \autoref{tab:accessCPU}. For the pull pattern in dense kernels, $3 Q$ memory accesses are required. 
One memory access is needed for the read of field A, one is needed for the write on field B, and the third one is a "write allocate B", which occurs if the data of the PDF of field B is not already stored in the CPU cache, and therefore has to be loaded into the cache to be written on. 
A PDF entry is only used once in a fused stream-collide step, and the whole \texttt{PDF-list} is unlikely to fit completely in the CPU cache for large-scale runs. Therefore, the "write allocate B" access is mostly present. This memory access only appears on CPUs, as GPUs do not utilize cache structure like CPUs do, so the amount of memory accesses on GPUs per cell is $2 Q$. Nevertheless, we want to avoid the third access on CPUs by utilizing an in-place streaming pattern. The data is already in the CPU cache because we read and write on the same PDF positions in the same PDF field. By this, we avoid the cache miss and end up with $2 Q$ memory accesses per cell.

For sparse kernels, utilizing the AA pattern has even more advantages in terms of memory access. 
For the pull pattern, in addition to the \texttt{PDF-list}, also the \texttt{index-list} has to be read to get the pull accesses for the propagation step, which adds a $(Q-1)$ to our count of memory accesses. 
In total, we need $3 \cdot Q + (Q-1)$ memory accesses for sparse LBM kernels with the pull streaming pattern. 

For the AA pattern on the other side, we only need neighboring information in every second (odd) time step because on even time steps, we only compute cell-local, and therefore no neighboring information is needed. 
So, the memory accesses for the index list can be halved to $(Q-1)/2$. 
Therefore, this results in $2 Q + (Q-1)/2$ memory accesses for sparse LBM kernels with the AA streaming pattern. 
We see in \autoref{tab:accessCPU} that the AA pattern on a CPU reduces the memory accesses compared to the pull pattern by
\begin{equation}
    1 - \frac{3 Q \cdot B_{\text{pdf}} + (Q-1) \cdot B_{\text{idx}}}{2 Q \cdot B_{\text{pdf}} + (Q-1)/2 \cdot B_{\text{idx}}}.
\end{equation}
Because well-optimized LBM codes are usually memory-bound, an increase in the performance of the LBM by the same ratio can be expected.

As already mentioned above, this performance boost can only be achieved on CPUs, as GPUs do not work with similar caches. No "write allocate" on the cache can be avoided. Therefore, only half the memory accesses for the \texttt{index-list} can be saved for the sparse approach, as shown in \autoref{tab:accessGPU}.

On the other hand, the condition to store the temporary PDF field can be avoided on CPUs and accelerators. 
So the memory consumption for the sparse LBM in \autoref{equ:mem} shrinks to 
\begin{equation}
    M_{\text{sparse,aa}} = N_{\text{cells}} \cdot (\underbrace{Q \cdot B_{\text{PDF}}}_\text{PDF-list} + \underbrace{(Q-1) \cdot B_{\text{idx}}}_\text{index-list} + \underbrace{5 \cdot B_{\text{PDF}}}_\text{other\ fields}) \cdot \phi.
\end{equation}
So for a D3Q19 stencil, double precision PDFs, and an integer \texttt{index-list}, we save 36.5\% of memory consumption by utilizing the AA streaming pattern for the sparse data structure.

\begin{table}
\centering
\caption{Memory accesses per cell for Pull and AA pattern on CPU with size of PDFs $B_{\text{pdf}} = 8$ Byte, size of indices in the \texttt{index-list} $B_{\text{idx}} = 4$ Byte and $Q = 19$.}
\begin{tabular}{ |c|c|c| } 
\hline
 \multicolumn{3}{|c|}{Memory accesses CPU} \\ \hline
             & Dense data & Sparse data \\ \hline
 Pull & $3 Q \cdot B_{\text{pdf}}$ &  $3 Q  \cdot B_{\text{pdf}} + (Q-1) \cdot B_{\text{idx}}$ \\ \hline
 AA & $2 Q \cdot B_{\text{pdf}}$ & $2 Q  \cdot B_{\text{pdf}} + (Q-1)/2  \cdot B_{\text{idx}}$ \\ \hline
  \begin{tabular}[x]{@{}c@{}}Reduction\\(1-AA/Pull)\end{tabular} & 33.3 \% & 35.6 \%  \\ \hline
\end{tabular}
\label{tab:accessCPU}
\end{table}

\begin{table}
\centering
\caption{Memory accesses per cell for Pull and AA pattern on GPU with size of PDFs $B_{\text{pdf}} = 8$ Byte, size of indices in the \texttt{index-list} $B_{\text{idx}} = 4$ Byte and $Q = 19$.}
\begin{tabular}{ |c|c|c| } 
\hline
 \multicolumn{3}{|c|}{Memory accesses GPU} \\ \hline  
              & Dense data & Sparse data \\ \hline
 Pull pattern & $2 Q \cdot B_{\text{pdf}}$ &  $2 Q  \cdot B_{\text{pdf}} + (Q-1) \cdot B_{\text{idx}}$ \\ \hline
 AA pattern & $2 Q \cdot B_{\text{pdf}}$ & $2 Q  \cdot B_{\text{pdf}} + (Q-1)/2  \cdot B_{\text{idx}}$ \\ \hline
 \begin{tabular}[x]{@{}c@{}}Reduction\\(1-AA/Pull)\end{tabular}  & 0 \% & 9.7 \%  \\ \hline
\end{tabular}

\label{tab:accessGPU}
\end{table}

\subsubsection{Benchmarking Results}
\begin{figure}[htb]
	\centering
    \includegraphics[width=\linewidth]{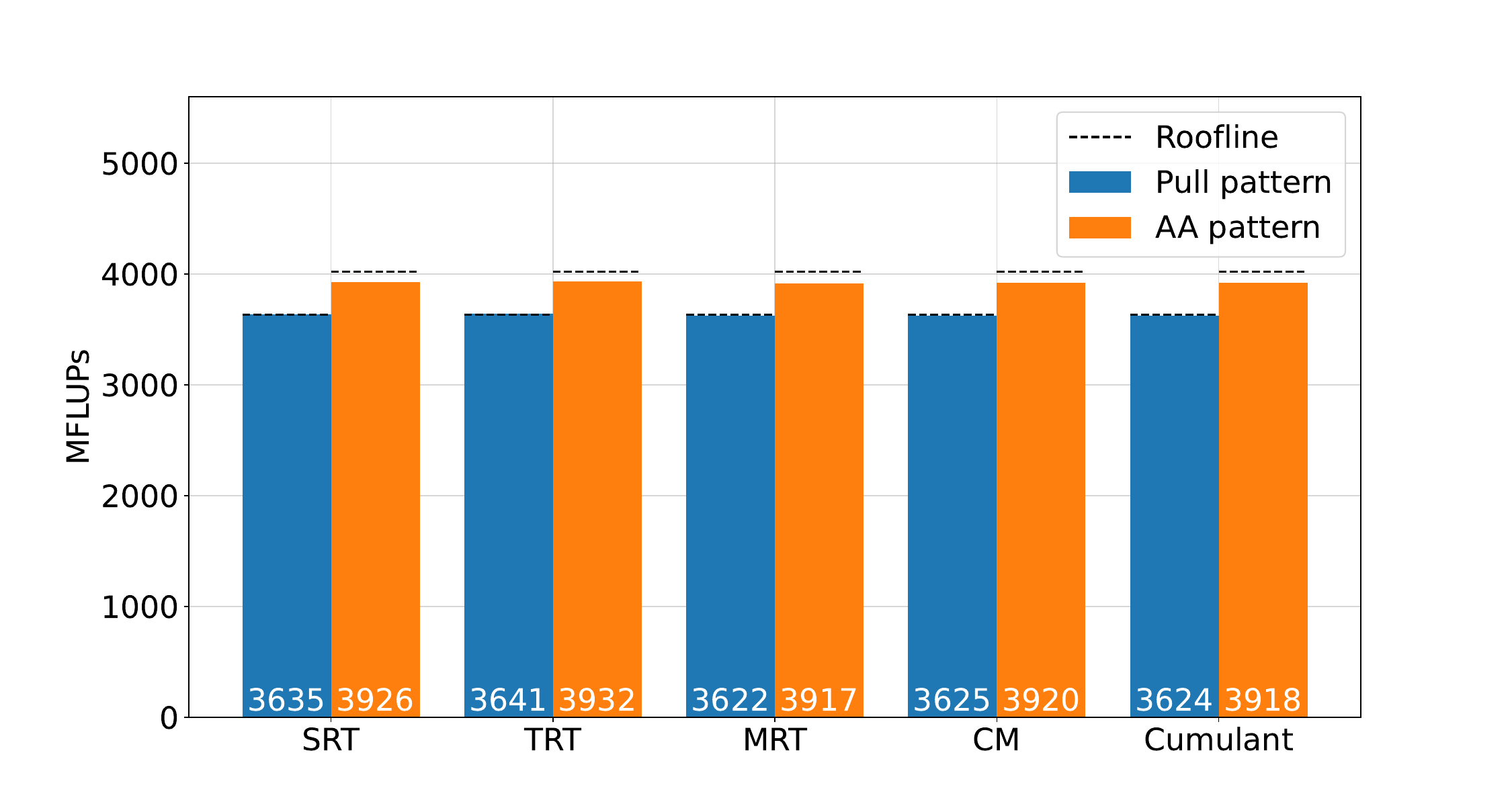}
    \caption{Single GPU Benchmark for pull vs AA streaming pattern for single relaxation time (SRT), two relaxation time (TRT), multi relaxation time (MRT), central moment (CM) and cumulant collision model on a D3Q19 stencil with $256^3$ cells on a single \nvidia{} A100.}
	\label{fig:CompPullAAGPU}
\end{figure}

In \autoref{fig:CompPullAAGPU} the single GPU benchmarking results for a sparse LBM kernel on a \nvidia{} A100 with a D3Q19 stencil and $256^3$ lattice cells are presented. 
We compare the pull streaming pattern with the AA pattern for various collision operators. 
The theoretical peak performance is calculated by the bandwidth, which is 1367 GB/s found by streaming benchmarks (\cite{siefertLatencyBandwidthMicrobenchmarks2023a}), and the number of theoretical memory accesses from \autoref{tab:accessGPU}.

We observe, that the performance of both streaming patterns is close to the theoretical performance for all of the presented collision operators. 
The average performance increase of the AA pattern compared to the pull pattern on accelerators is $\sim 7.5 \%$, which is close to the theoretically achievable performance increase of $9.7 \%$ from \autoref{tab:accessGPU}. 
Therefore, in addition to the avoidance of the storage of the second PDF field, it is also worth to employ the AA pattern on GPUs in terms of performance.

\subsection{Communication Hiding}
\begin{figure}[htb]
	\centering
	\includegraphics[width=0.5\linewidth]{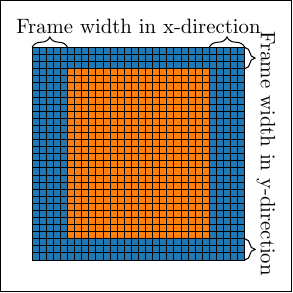}
    \caption{Subdivision of the PDF field in a frame and the block interior to enable communication hiding. In this example, the frame width is 5 in x, and 3 in y-direction.}
	\label{fig:CommHiding}
\end{figure}

Communication hiding is used to overlap communication with computation. 
For this, the domain on every block has to be divided into a "block interior" and a "frame", as illustrated in \autoref{fig:CommHiding}. The frame only consists of the outermost cells, while the block interior consists of all other cells.
An exemplary code to achieve communication hiding is shown in Algorithm \autoref{algo:commHiding}.
At first, the communication is started. 
Every block packs its outermost PDFs in an MPI buffer and performs a non-blocking MPI-Send to its neighbors. 
Now, the block interior cells can be updated because the information from neighbour blocks is not needed for these cells. 
After this step, the algorithm must wait for the communication to complete and write the information of the MPI buffers to the ghost layers. 
Lastly, with the updated information in the ghost layers, the LBM and boundary kernels can now be executed on the cells of the frame. 

With this algorithm, the communication of the simulation can be overlapped with the kernel on the interior, which leads to higher performance because of better scalability on an increasing number of MPI processes. 
The width of the frame in all three dimensions has to be chosen suitably to achieve best possible performance. 
A thinner frame width would increase the number of cells in the interior, providing more time to overlap the communication. 
On the other hand, a small frame width results in small kernels. Especially on GPUs, small kernels can not fully utilize the GPU, which can lead to performance drops.
Additionally, consecutive memory access in one dimension is not possible for a thin frame, which can also reduce the simulation's performance.

\begin{algorithm}
\caption{Communication Hiding}\label{algo:commHiding}
\begin{algorithmic}[1]
\For {each time step}
\State Start communication
\State
\State Run boundary kernels on block interior
\State Run LBM kernel on block interior
\State
\State Wait communication
\State
\State Run boundary kernels on frame
\State Run LBM kernel on frame
\EndFor
\end{algorithmic}
\end{algorithm}

\subsubsection{Communication Hiding for Sparse Data Structures}
Implementing communication hiding for a sparse data structure is not straightforward because there is no spatial information for the cells. 
This means that a cell has no direct information about whether it is inside the block interior or part of the frame. 
To compensate this, we store two additional index lists, one for the interior and one for the PDFs on the frame. 
These index lists are initialized by the flag field at the start of the simulation, where spacial information of cells is still present. 
Furthermore, the pull index for the center PDF of every fluid cell must be stored. This index is used to get the correct write access for kernels on the interior and frame cells. 
These modifications on the list structure are integrated into the code generation so that they can be turned on and off and allow generated code to run on different architectures.

\subsubsection{Scaling results}
\begin{figure}[htb]
	\centering
     \includegraphics[width=\linewidth]{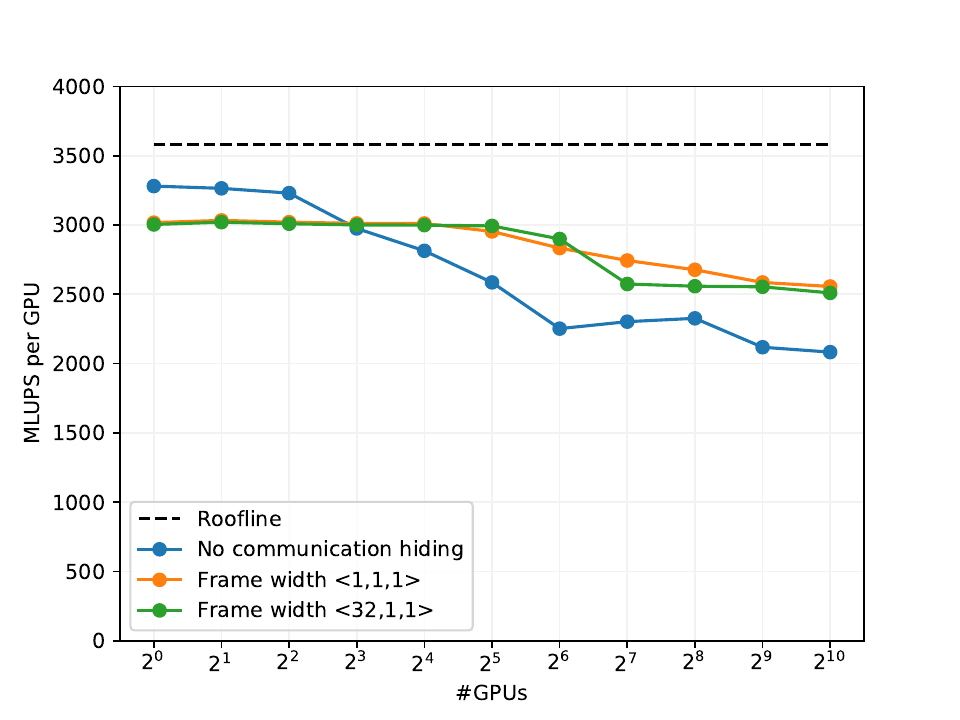}
    \caption{Weak scaling benchmark on \nvidia{} A100 GPU cluster JUWELS Booster with different configurations for the communication hiding. The roofline is obtained by a stream benchmark (\cite{siefertLatencyBandwidthMicrobenchmarks2023a}). The runs are executed with $320^3$ cells per GPU, with a D3Q19 stencil and SRT collision model on an empty channel setup.} 
	\label{fig:BenchmarkCommHiding}
\end{figure}

In \autoref{fig:BenchmarkCommHiding}, the weak scaling of the sparse data structure on the JUWELS Booster HPC cluster is presented. JUWELS Booster is currently place 21 of the Top500 HPC systems (June 2024) and consists of 936 compute nodes, each equipped with 4 \nvidia{} A100 GPUs (see \cite{juwels}). 

We tested three versions of the communication. 
One is without communication hiding, one with the minimum frame size of one cell in every direction, and one with a frame thickness of 32 cells in x direction and one cell in y and z direction. 
This option is promising, since consecutive memory accesses are still enabled in x-direction, while the frame size is still small enough to allow a good communication overlap.
In general, a smaller frame size increases the work of the kernels on the interior cells and, therefore, should increase the effectiveness of the communication hiding. On the other hand,  the GPU utilization of the kernels on the frame is quite low for a small frame size, and consecutive memory accesses are not secured.

The version without communication hiding performs best up to one node (4 GPUs), see \autoref{fig:BenchmarkCommHiding}. There the intra-node communication speed is quite high since we can exploit the high bandwidth of \nvidia{} GPU-to-GPU connections. However, the performance without communication deteriorates for more than 4 GPUs when the inter-node communication speed becomes relevent. 

The benchmark runs with communication hiding start with worse performance on single node, as the overhead of the kernel call on the frame cells limits the performance. 
Nevertheless, these versions exhibit excellent scalability for up to 32 GPUs. 
Beyond 32 GPUs, the performance drops to 83\% scaling efficiency on 1024 GPUs. 
This can possibly be explained by the InfiniBand network architecture of JUWELS Booster, which is implemented as a DragonFly+ network. 
The drop in performance could be caused by the need for communication between different switch islands of the system when more than 32 GPUs are employed. 

In these cases, the size of the frame only has a negligible impact. The two scenarios for communication hiding behave similarly. The smaller frame size of $<1,1,1>$ performs a bit better on more than 128 GPUs.
Nevertheless, we see that communication hiding can increase the scaling efficiency of the sparse data structure on up to 1024 \nvidia{} A100 GPUs from 63\% to 83\%.

\begin{figure}[htb]
	\centering
     \includegraphics[width=\linewidth]{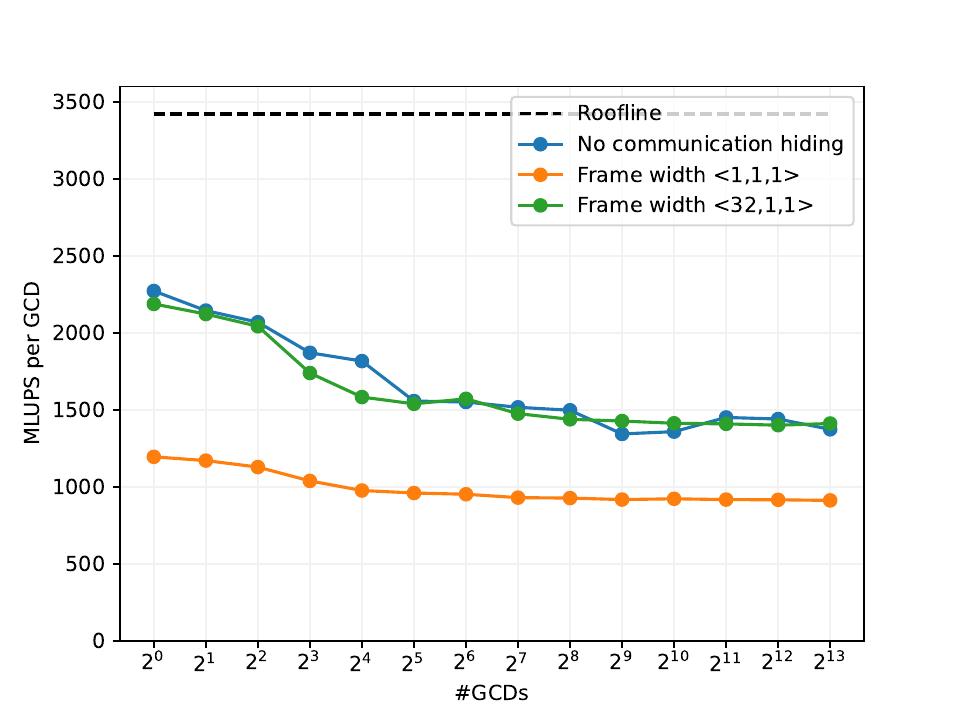}
    \caption{Weak scaling benchmark on GPU cluster LUMI-G with different configurations for the communication hiding. The roofline is obtained by a stream benchmark (\cite{siefertLatencyBandwidthMicrobenchmarks2023a}) . The \amd{} MI250X GPUs have two compute chips per GPU (GCDs). The runs are executed with $256^3$ cells per GCD, with a D3Q19 stencil and SRT collision model on an empty channel setup.}
	\label{fig:BenchmarkCommHidingLUMIG}
\end{figure}

Additionally, we tested the scaling efficiency on the GPU partition of LUMI, which is in the top 5 on the current Top500 list (June 2024). The HPC cluster comprises 2978 nodes with 4 \amd MI250X GPUs per node. 
Further, every \amd{} MI250X GPU consists of two Graphical Compute Dices (GCDs) (\cite{pearsonInterconnectBandwidthHeterogeneity2023}), so we create one MPI process per GCD and show the scaling over the GDCs. 
As already studied in \cite{holzerDevelopmentCentralmomentPhasefield2024}, \cite{lehmannEsotericPullEsoteric2022} and \cite{martinPerformanceEvaluationHeterogeneous2023}, it seems not to be possible to achieve significantly better performance than equivalent to approximately 50\% of memory bandwidth for LBM codes on a single \amd MI250X. We observe the same behavior in \autoref{fig:BenchmarkCommHidingLUMIG}.

We tested the three communication routines similar to the benchmarks on JUWELS Booster. 
For the runs without communication hiding, the scaling behavior is similar as for the larger frame size of $<32,1,1>$. 
For the cases without communication hiding, we achieve a CDG scaling efficiency of 60\% scaling from one to 8192 CDGs (4096 GPUs) and a node (4 GPUs) scaling efficiency of 82\%. 
For the greater frame size $<32,1,1>$ we observe similar scaling behavior as without communication hiding. 
The expected acceleration and better scaling could not be observed, a finding that should be further investigated in future research. 
For the small frame size, the scaling efficiency is almost perfect, but the overall performance is much worse than for the other communication strategies. 
Again this behavior is unexpected. The non-consecutive memory accesses and kernel calls with small execution times could be the reason for the relatively pour overall performance of the simulation in these cases.

\section{Applications}

To evaluate the performance of the sparse data structure in a more realistic scenario than the artificial porosity benchmark as in \autoref{fig:SparseVsDense} or the weak scaling of an empty channel on JUWELS Booster (\autoref{fig:BenchmarkCommHiding}) and on LUMI (\autoref{fig:BenchmarkCommHidingLUMIG}), we set up three different applications. 
The first is a flow through a porous medium consisting of a stationary particle bed. 
The second one is an extended version of the first application, where the bottom part of the domain consists of the same particle bed, while the upper domain is a free flow, such that we simulate the interaction of a free flow with a porous sediment bed. 
The last application is the flow through a geometry of coronary arteries, which also results in a complex and sparse domain. 

\subsection{Flow through Porous Media}
\begin{figure}[htb]
	\centering
    \includegraphics[width=\linewidth]{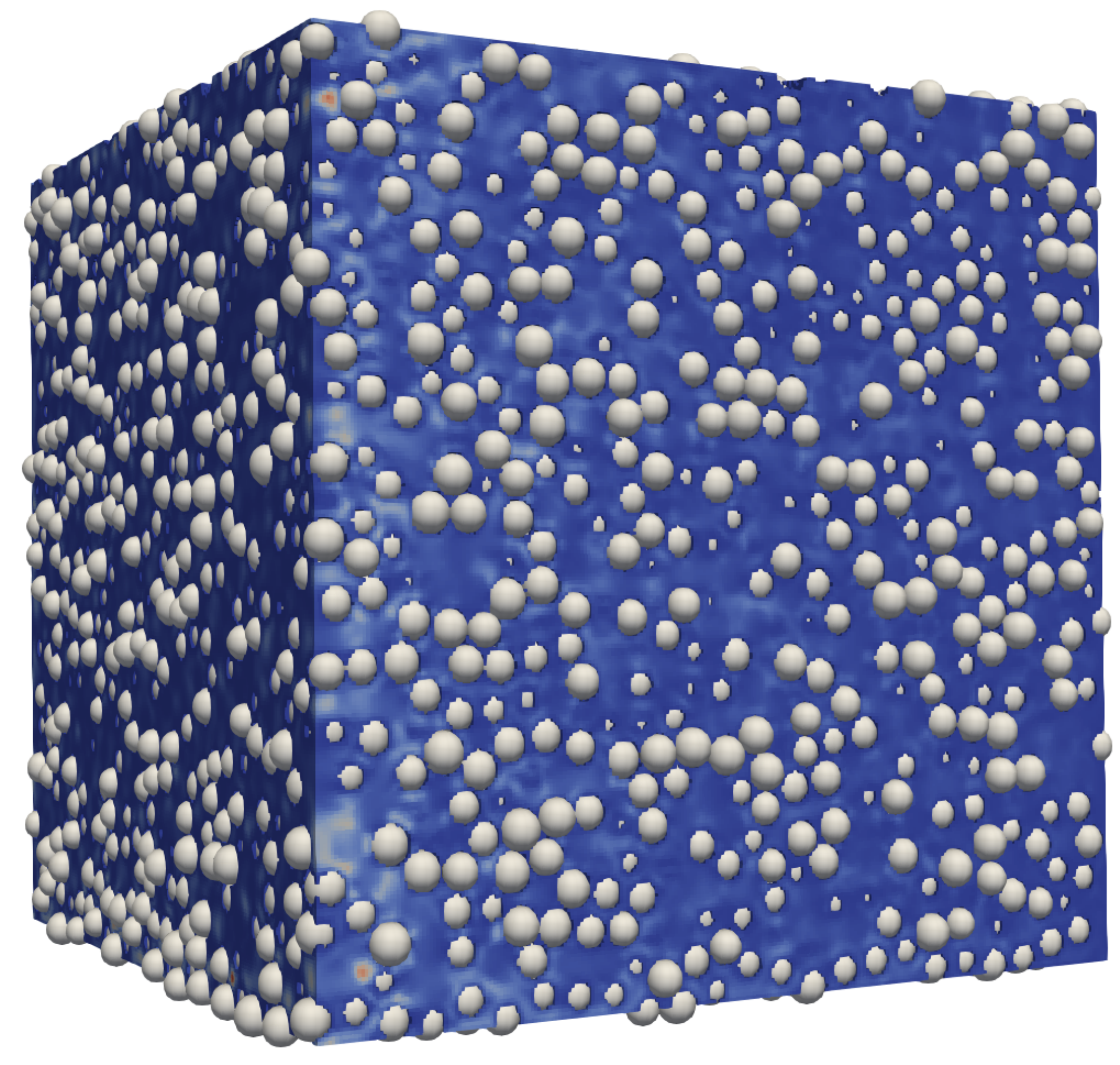}
    \caption{Flow through a particle bed consisting of 21580 particles of 0.0041m diameter, which corresponds to an average porosity of 0.356663. The size of the domain is 0.1m in every dimension.}
	\label{fig:ParticleBed}
\end{figure}

The efficient simulation of fluid flow through porous media is an ongoing research topic, for example in \cite{panHighperformanceLatticeBoltzmann2004}, \cite{yangImplementationDirectaddressingBased2023}, \cite{hanLBMModelingFluid2013} or \cite{ambekarParticleResolvedTurbulent2023a}, to mention a few.
For this porous application, we generated a particle bed with the \walberla{} molecular dynamics module MESA-PD, as shown in \cite{rettingerCoupledLatticeBoltzmann2018}. 
We defined a domain of 0.1 meters in every dimension and filled it with 21580 particles with a diameter of 0.0041m. This setup is illustrated in \autoref{fig:ParticleBed} and results in an average porosity of 0.356663. 

\begin{figure}[htb]
	\centering
    \includegraphics[width=\linewidth]{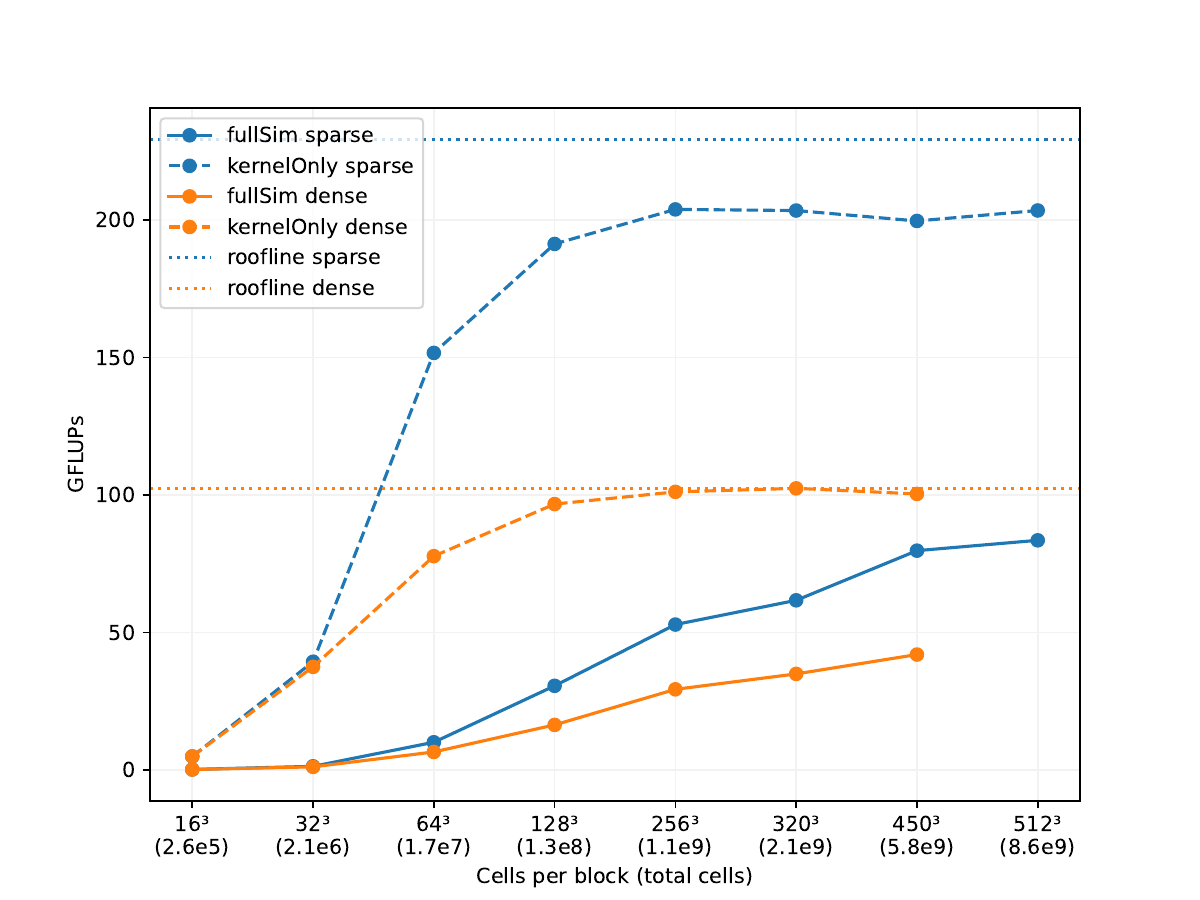}
    \caption{Comparison of the sparse and the dense data structure for the flow through the particle bed in \autoref{fig:ParticleBed}. The number of \walberla{} blocks is fixed to 64 while the cells per blocks increases, and therefore also the resolution and the number of total cells increases. The benchmark was executed on 64 \nvidia{} A100 GPUs on JUWELS Booster with one block per GPU.}
	\label{fig:ParticleBedSparseVsDense}
\end{figure}

We decomposed the domain with 64 blocks in a 4x4x4 arrangement to run on 64 \nvidia{} A100 GPUs on JUWELS Booster. 
While we fixed the number of blocks to 64, we increase the cells per block and therefore also the resolution of the domain and the number of total cells, as shown in \autoref{fig:ParticleBedSparseVsDense}. 
We performed this benchmark for the sparse data structure, and compare it to the dense data structure. 

We first focus on at the "kernel-only" results in \autoref{fig:ParticleBedSparseVsDense}, which only run the LBM kernel without handling boundaries or performing communication. We observe, that for block sizes below $64^3$, the GPU utilization is too low to achieve good performance. 
Both, the sparse and the dense kernel saturate at a block size of $256^3$. 
Further, we note that the performance of the sparse data structure is approximately two times higher than for the dense structure. 
This also fits quite well to the results in \autoref{fig:SparseVsDense} for a porosity of $\phi \sim 0.35$. 
The sparse kernel-only performance does not quite reach the theoretical bandwidth limit. This could be caused by some in-balances, since the porosity varies between the blocks, from a minimum of 0.337816 to a maximum of 0.392441. 
This means, that some blocks, and therefore processes, have more workload in terms of cells.
We measure the performance at the end of the simulation run, when all processes finished their work, so the performance is determined by the slowest processor. 
The performance of the dense kernel on the other side is not affected by the porosity differences, and therefore performs exactly the same work on every MPI process and thus does not suffer from load in-balance issues.

When running the full simulation including boundary handling and communication routines, we again observe low performance for small block sizes, which can be explained by the low utilization of GPUs. 
However, this time, there is no saturation for a block size of $256^3$ because a more extensive block size results in better communication hiding when the ratio between computational work and communication improves in favor of computational work. 
For the dense structure, it was not possible to perform simulations with a block size of $512^3$, as the memory consumption exceeded the 40 GB of the GPU RAM of a single A100. 
For the sparse structure, this is not a problem, as for a porosity of 0.356663 we save around 50\% of memory compared to the dense structure. 

For this application, we significantly benefit from the sparse data structure, as it achieves a performance increase of $\sim 90\%$ compared to the dense one for a block size of $450^3$. 
For a block size of $512^3$, which results in a cell resolution of $4.8828*10^{-5}$m and $8.6*10^9$ total cells, it achieves an overall performance of 203.408 GFLUPs for the kernel-only call and 83.475 GFLUPs for the full-simulation run. 
Further, as shown in \autoref{fig:SparseVsDenseMemory}, for a domain with an average porosity of $\sim 0.35$, we are able to save around 50\% of memory consumption by utilizing the sparse data structure.

\subsection{Free Flow over River Bed}

\begin{figure}[htb]
	\centering
    \includegraphics[width=\linewidth]{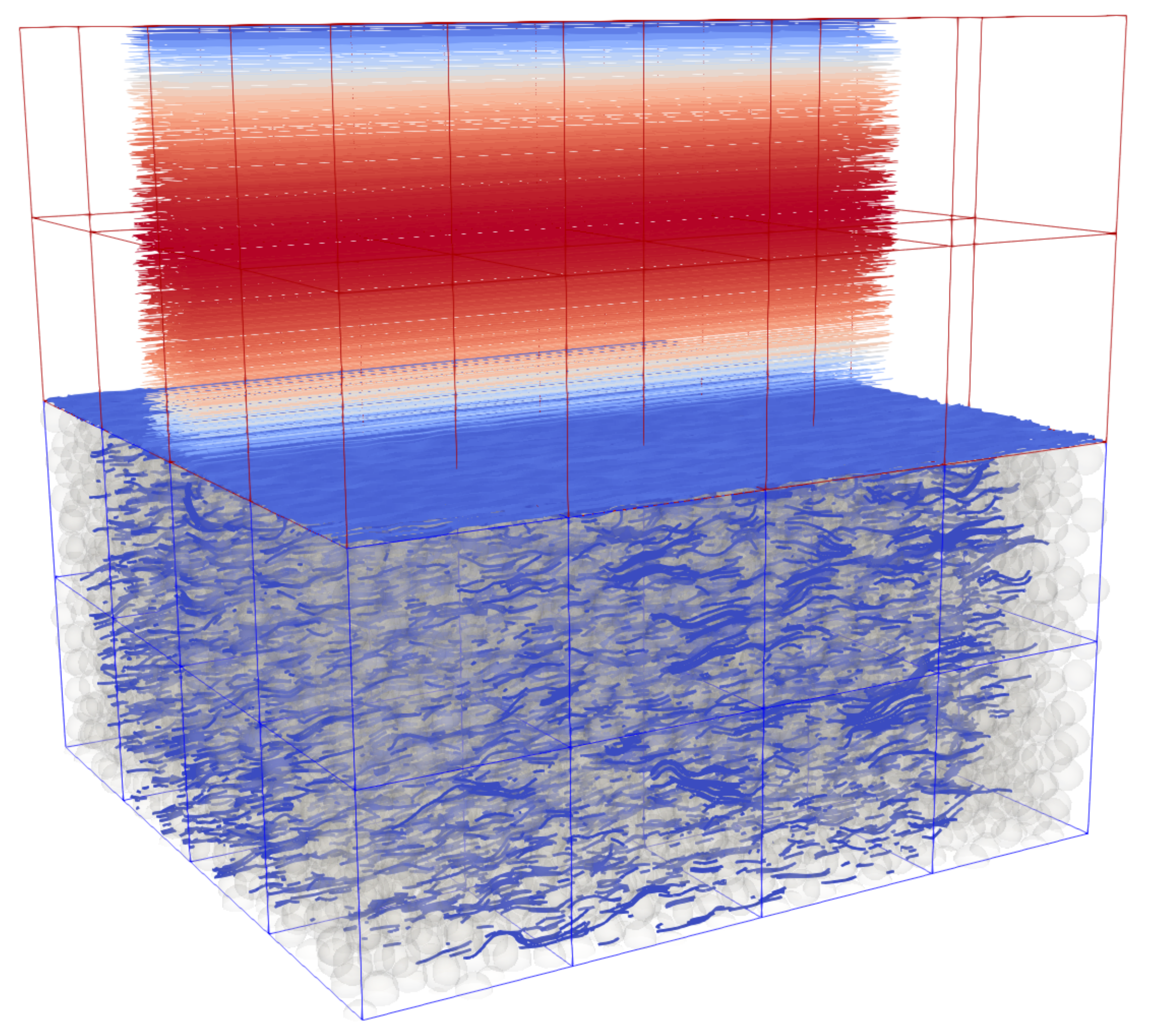}
    \caption{Free Flow over a particle bed. The porosity of the blocks in the particle bed on the bottom (blue blocks) have a porosity of about 0.35, while the upper blocks (red blocks) consists of fluid cells only.}
	\label{fig:ParticleBedFreeFlow}
\end{figure}

The second application is the simulation of a free flow over a river bed similar to \cite{antidunes} or \cite{fattahiLargeScaleLattice2016}. 
In \autoref{fig:ParticleBedFreeFlow}, the bottom part of the domain consists of the same porous medium / particle bed as the first application in \autoref{fig:ParticleBed}, with the same average porosity of about $\sim 0.35$. The upper part of the domain is a free flow only, so 100\% fluid cells in this part of the domain. The average porosity of the domain is $\sim 0.68$. 

This application is suitable for utilizing the hybrid data structure. 
As already indicated in \autoref{fig:ParticleBedFreeFlow}, the blocks in the upper part of the domain should hold their data in a dense structure (red blocks). 
In contrast, the blocks in the porous part of the domain should be stored with a sparse data structure (blue blocks). 
The framework includes the functionality to select the appropriate data structure for each block, the user only has to specify an appropriate porosity threshold.

\begin{figure}[htb]
	\centering
    \includegraphics[width=\linewidth]{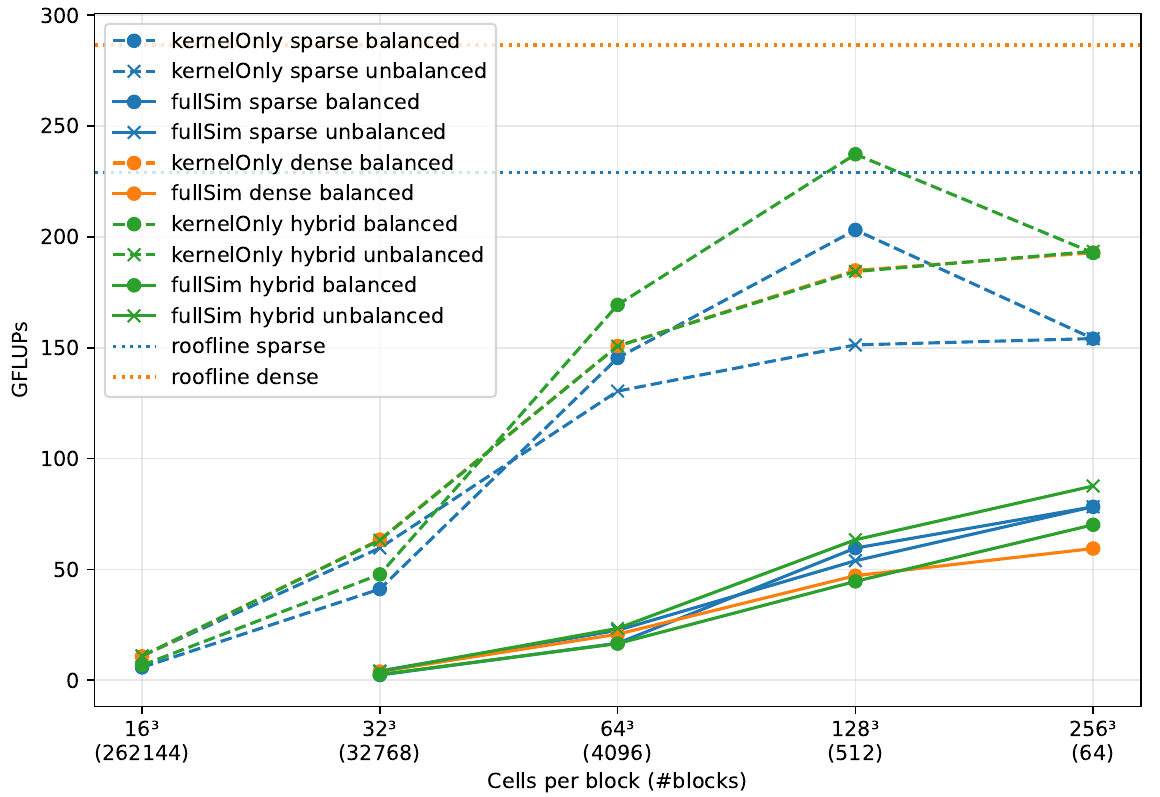}
    \caption{Comparison of the sparse, dense and hybrid data structure for the free flow over a particle bed in \autoref{fig:ParticleBedFreeFlow} on 64 \nvidia{} A100 GPUs on Juwels Booster. 
    The problem size is fixed to $1.07 * 10^9$ cells, while the number of cells per block and therefore also the number of blocks vary.
    The LBM kernel-only performance is shown as well as the performance of the whole simulation including boundary handling and communication between MPI processes.}
	\label{fig:ParticleBedFreeFlowSparseVsDense}
\end{figure}

In \autoref{fig:ParticleBedFreeFlowSparseVsDense}, a comparison of the sparse, the dense and the hybrid data structure is shown. 
The simulations are again executed on the JUWELS Booster GPU cluster, with a fixed number of \nvidia{} A100 GPUs and one MPI process per GPU. 
The performance of the raw  LBM kernel (kernel-only) is plotted, as well as the entire simulation run, including communication between the MPI processes. 
We fixed the number of GPUs to 64 as well as the problem size to $10^9$ cells and only vary the cells per block, resulting in a high number of blocks for small cells per block and one block per GPU for the largest block size of $256^3$ cells.

Evaluating the raw kernel performance first, we again observe that a more significant number of cells per block leads to a better utilization of the GPU. 

Load-balancing issues can emerge when using a sparse or a hybrid data structure with multiple blocks per GPU. 
This is because the block partitioning of a domain can lead to a wide span of porosity values on the blocks. 
When decomposing the river bed simulation in \autoref{fig:ParticleBedFreeFlow} with sparse blocks only, this results in half of the blocks yielding a porosity of $\sim 0.35$ and the other half yielding a porosity of $1.0$. 
One can calculate the workload of a sparse \walberla{} block on a GPU similar to the number of memory accesses in \autoref{tab:accessGPU} with 
\begin{equation}
    w_{\text{sparse}} = (\underbrace{2 \cdot Q \cdot B_{\text{PDF}} }_{\text{PDF list read/write}} + \underbrace{(Q-1) \cdot B_{\text{idx}}}_{\text{Index list read}}) \cdot \phi
\end{equation}
with Q as the stencil size and $\phi$ as the porosity. 

The workload for the dense block on a GPU is 
\begin{equation}
        w_{\text{dense}} = \underbrace{2 \cdot Q \cdot B_{\text{PDF}} }_{\text{PDF field read/write}}, 
\end{equation}
which is not depending on the porosity of the block. 

For the simulation run with sparse data blocks only, the workload of the blocks differs significantly depending on their porosity. 
Therefore, we employ a load-balancing algorithm to balance the blocks over the MPI processes to reach a better workload distribution. 
We used a space-filling-Hilbert-curve approach, as described in \cite{schornbaumExtremeScaleBlockStructuredAdaptive2018}.

In \autoref{fig:ParticleBedFreeFlowSparseVsDense}, we observe that the load-balancing works well for the sparse kernel-only runs. Especially for the block sizes of $64^3$ and $128^3$, the load-balancing seems to reduce the workload unbalance significantly and, therefore, increases the performance. For $256^3$ cells per block, there is only one block per GPU, so no load-balancing is possible in this case. 

No load balancing is necessary for the dense data structure, as every block has the same workload. We see that the kernel-only runs of the dense data structure outperform the unbalanced sparse runs. 
This is because the sparse data structure is slower than the dense structure on the blocks with a porosity of 1.0 (see again \autoref{fig:SparseVsDense}), and all MPI processes must wait for the processes holding these blocks. 
However, when balancing the workload for the sparse kernels, at least for a block size of $128^3$, we can achieve a higher performance than the dense kernels. 

The kernel-only runs for the hybrid data structure without load balancing closely follow the results of the dense kernels in \autoref{fig:ParticleBedFreeFlowSparseVsDense}. 
Without load balancing, the low performance of the dense blocks in the porous region heavily dominates the runtime. 
However, here we can balance the workload to achieve better results. 
So for the kernel-only runs, using the workload-balanced hybrid data outperforms the other kernels for all block sizes. The only exception is for $256^3$ cells per block. There no load-balancing is possible, because there is only one block per GPU. 

We observe a similar behavior when studying the entire simulation runs in \autoref{fig:ParticleBedFreeFlowSparseVsDense}. 
This now includes the communication between the MPI processes. Overall, the sparse data structure performs superior to the dense one on the block sizes of $128^3$ and $256^3$ cells per block. 

For the hybrid structure, we observe unexpected behavior. 
Here, the load-balanced simulation performs worse than the sparse simulations, while the unbalanced hybrid simulation reaches the highest MFLUPs values of all tested data structures. In \autoref{tab:workload}, the workload per MPI process of the simulation with the hybrid data structure for $128^3$ cells per block is presented. 
We observe that the standard deviation of the average workload is significantly lower than for the unbalanced execution. 
Still, the performance of the balanced run is worse for the full-simulation. 
Here we observe that the load-balancing algorithm successfully distributes the workload over the MPI processes but it reduces the spatial locality of the blocks with respect to each other. 
Therefore, the communication is more expensive. 

Nevertheless, for a complex domain setup such as a free flow over a riverbed, we also manage to improve the application's performance by utilizing the sparse or the hybrid data structure. 
Additionally, we note the reduced memory consumption as presented in \autoref{fig:SparseVsDenseMemory}. 
Overall, when half of the domain consists of a porous medium, such as in \autoref{fig:ParticleBedFreeFlow}, we save about 25\% of memory with the hybrid data structure.

\begin{table}
\centering
\caption{Workload per MPI process / GPU before and after load balancing for the hybrid data structure run with $128^3$ cells per block in \autoref{fig:ParticleBedFreeFlowSparseVsDense}.}
\begin{tabular}{ |c|c|c|c|c| } 
\hline
 Workload & Average & Std deviation & Min & Max \\ \hline
 Unbalanced & 1757.47 & 675.194 & 1038 & 2432 \\ \hline
 Balanced & 1757.47 & 102.644 & 1520 & 1826 \\ \hline
\end{tabular}
\label{tab:workload}
\end{table}

\subsection{Coronary Artery}
\begin{figure}[htb]
	\centering
    \includegraphics[width=\linewidth]{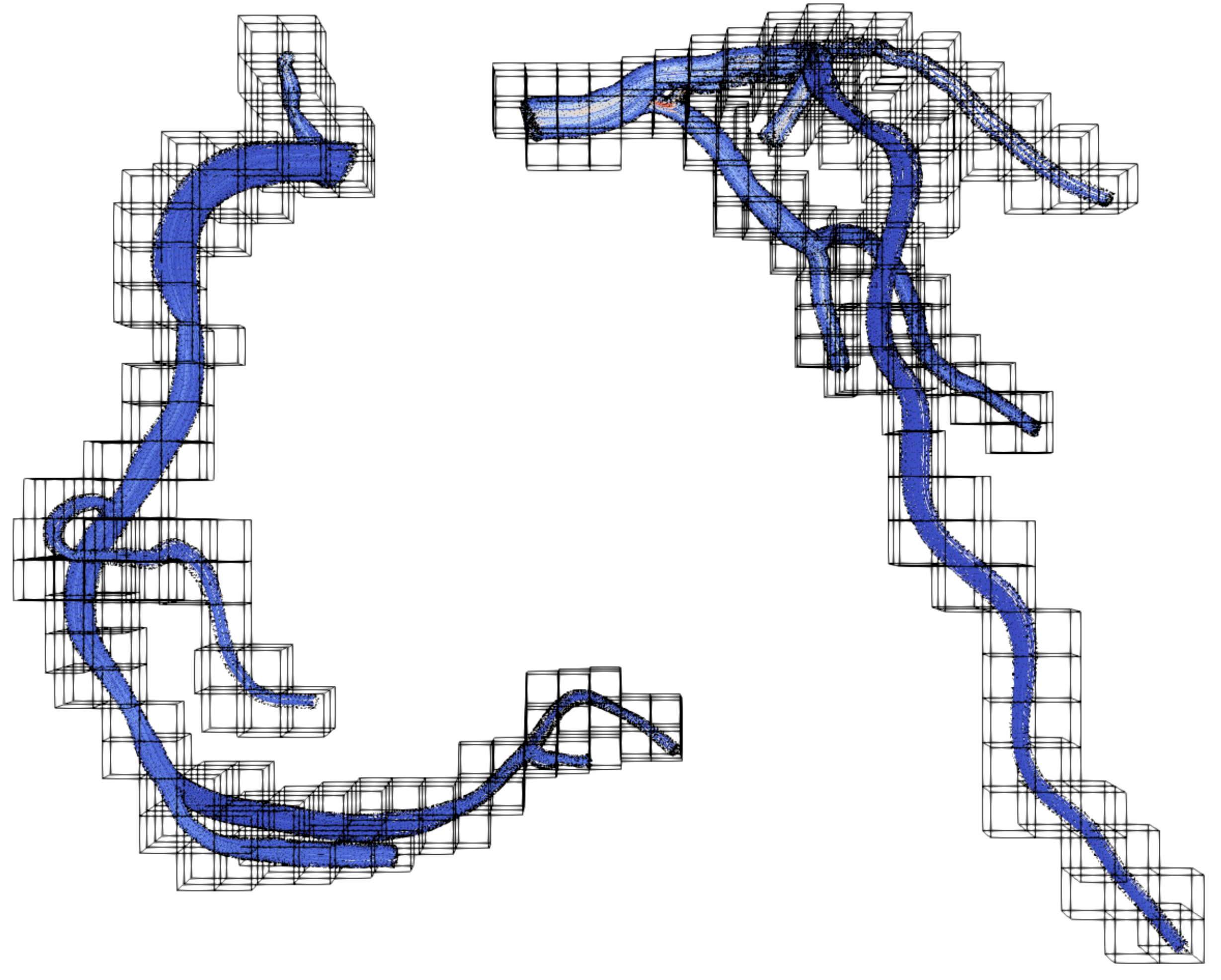}
    \caption{Domain partitioning of a coronary artery with  $1.2 \cdot 10^8$ fluid cells, 531 blocks and $128^3$ cells per block.}
	\label{fig:Artery}
\end{figure}

Finally, we present performance results for the flow in a coronary artery. 
This is a topic of high interest in medical engineering and has been studied, for example, in \cite{axnerSimulationsTimeHarmonic2009}, \cite{afrouziSimulationBloodFlow2020a}, \cite{bernaschiFlexibleHighPerformance2010} and \cite{godenschwagerFrameworkHybridParallel2013}.
The flow in a coronary artery results in a complex and highly sparse domain. 
An example setup for this application is shown in \autoref{fig:Artery}. 
While the whole domain would consist of a very high number of blocks, we can discard all blocks that do not contain fluid cells. Still, the remaining blocks have a very low porosity. 
Of course, when lowering the size of the block, the block structure would converge better to the geometry and result in higher block porosities. However, small block sizes also lead to under-utilization of GPUs, as already shown in \autoref{fig:ParticleBedSparseVsDense}. 

\begin{figure}[htb]
	\centering
    \includegraphics[width=\linewidth]{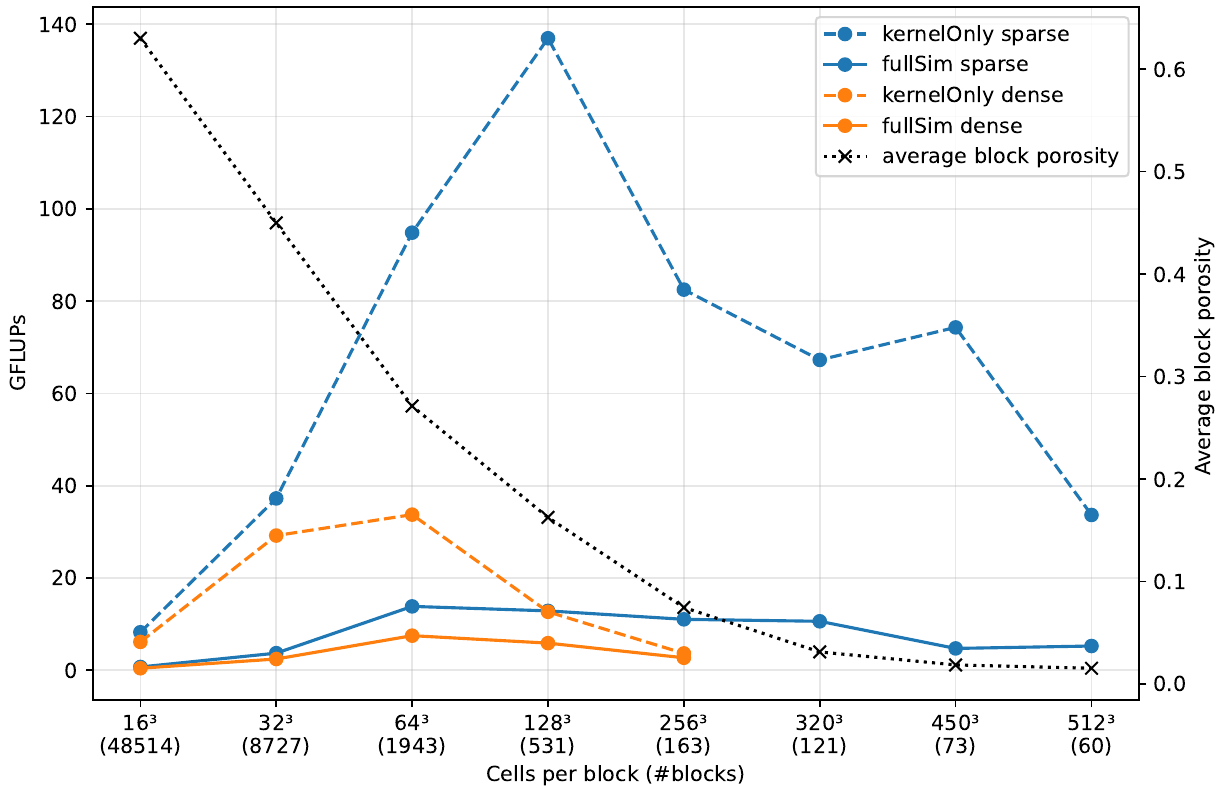}
    \caption{Comparison of the sparse and the dense data structure for the artery flow in \autoref{fig:Artery} on 60 \nvidia{} A100 GPUs on JUWELS Booster. The problem size is fixed to a number of $1.2 \cdot 10^8$ fluid cells, while the block size and therefore also the number of blocks varies.}
	\label{fig:ArterySparseVsDense}
\end{figure}

In \autoref{fig:ArterySparseVsDense}, we compare the performance of the sparse and the dense data structure on JUWELS Booster. We fixed the number of \nvidia{} A100 GPUs to 60 and the problem size to $1.2 \cdot 10^8$ fluid cells while varying the block size and, therefore, also the number of blocks.
Again, we observe the behavior of a small block size, which results in low performance for the kernel-only runs of the sparse data structure. We experience that increasing the block size up to $128^3$ cells per block leads to higher performance of the raw sparse LBM kernel. 
For block sizes larger than $128^3$, the porosity drops below $\phi < 0.1$. 
We observe in \autoref{fig:SparseVsDense} that the sparse LBM performance starts to deteriorate for a porosity smaller than $0.1$. 
Therefore, for block sizes larger $128^3$, the performance of the sparse kernel in \autoref{fig:ArterySparseVsDense} shrinks because of the low block porosities. The sweet spot between large block sizes for good GPU utilization and a porosity greater than $0.1$ for good kernel performance is a block size of $128^3$ cells per block for this setup. 
The same behavior does not emerge for the sparse full simulations. 
Here, the performance is relatively stable for moderate block sizes. 
This is because the performance is heavily dominated by communication, which is especially high in this application because of the high number of blocks per MPI process. 

Nevertheless, the sparse data structure shows significantly higher performance than the dense data structure for all block sizes. This is true for the kernel-only runs and the full simulation runs. For a block size of $128^3$ cells, we achieve a speed-up of about 11 for the kernel-only run and still a speed up of 2 for the full simulation.

Also remarkable is the amount of memory we can save for the artery setup. 
So when choosing a block size of $128^3$ cells per block to reach maximum raw performance; then, we end up with an average block porosity of $\phi \sim 0.16$. This means that according to \autoref{fig:SparseVsDenseMemory}, we save about 75\% of memory when using the sparse instead of the dense data structure. 
We were not able to acquire results for the dense structure for block sizes greater than $256^3$ because the \nvidia{} A100 RAM ran out of memory.

Parts of this work are comparable to \cite{martinPerformanceEvaluationHeterogeneous2023}. In this article, the framework HARVEY is ported to different programming models such as CUDA, HIP, SYCL and Kokkos. The authors compare the performance on different hardware, among others also on NVIDIA A100 and AMD MI250X GPUs. HARVEY is a LBM based CFD software with a focus on simulations of blood flow in patient-derived aortas, also utilizing the indirect addressing approach.

\cite{martinPerformanceEvaluationHeterogeneous2023} measure the performance of HARVEY and a proxy app for NVIDIA A100 GPUs on the HPC System Polaris\endnote{Polaris: \url{https://www.alcf.anl.gov/polaris}, accessed: 2024-8-2}. 
Their proxy app is used to show the maximum achievable performance for simple test cases.
Their results of the proxy app for an empty channel flow on one node (4 NVIDIA A100 GPUs) reaches around 12000 MLUPS, equivalent to 3000 MLUPS per A100 GPU. 
This is close to what we achieve, as shown in \autoref{fig:BenchmarkCommHiding} for the communication-hiding cases. The actual HARVEY framework performs slightly worse for the same empty channel with around 2250 MLUPS per GPU. 
On the maximum scaling size of 1024 NVIDIA A100 GPUs, the proxy-app reaches around $10^4$ MLUPS, equivalent to 976 MLUPS per GPU. The HARVEY framework achieves around $6*10^4$ MLUPS, equivalent to 585 MLUPS per A100 GPU. In \autoref{fig:BenchmarkCommHiding} \walberla{} achieves 2500 MLUPS per GPU on 1024 NVIDIA A100 GPUs. We conclude that the scaling efficiency of \walberla{} seems to be superior. The results may not be fully comparable, as the team of \cite{martinPerformanceEvaluationHeterogeneous2023} performs a piece-wise strong-scaling, while we show the weak scaling of our software. Further, they operated on a different HPC cluster, so especially the scaling performance can depend significantly on the configuration of the HPC system. 

When comparing the results for the artery geometry with a resolution of 55 $\mu$m, HARVEY achieves around $4*10^4$ MLUPS for 64 A100 GPUs, so 625 MLUPS per GPU. In \autoref{fig:ArterySparseVsDense} \walberla{} was able to achieve 1374 MLUPS per GPU for the kernel-only runs, but only 230 MLUPS for the full simulation performance. Of course, the underlying geometry of the artery is different, so again the results may not be fully comparable. Nevertheless, there is still room to optimize the \walberla{} framework in terms of load-balancing and communication-efficiency, especially for cases with multiple blocks per GPU.

Further, \cite{martinPerformanceEvaluationHeterogeneous2023} show comparable results on the AMD MI250X GPUs, which were tested on the Frontier\endnote{Frontier: \url{https://www.olcf.ornl.gov/olcf-resources/compute-systems/frontier/}, accessed: 2024-8-2} HPC system. 
We compare the results for 4 GCDs. 
The proxy app of shows around 2250 MLUPS per GCD, while \walberla{} shows 2100 MLUPS per GCD (see \autoref{fig:BenchmarkCommHidingLUMIG}). So also the team of \cite{martinPerformanceEvaluationHeterogeneous2023} confirm, that LBM algorithms are not able to achieve more than 60\% of the maximum theoretical performance of the AMD MI250X GPUs. 
For 1024 CDGs, the proxy app shows a performance of around $10^6$ MLUPS, so 976 MLUPS per CDG. 
\walberla{} was able to achieve 1500 MLUPS per GCD on the same number of GCDs, so it shows a slightly better scaling efficiency. 
Again, the results are not fully comparable, as we compare a weak scaling with a piece-wise strong-scaling and also run on different systems.

\section{Conclusion}

In this article, we presented the benefit of sparse LBM kernels, especially when using accelerator cards. 
We compared the sparse and the dense data structure on a single GPU and found that for a domain porosity of $< 0.8$, the sparse data structure outperforms the dense data structure in terms of performance and memory consumption.  

The sparse kernels show excellent performance for the pull streaming pattern for various collision operators such as single-/two-/multi-relaxation times, central moments, or cumulants on a single GPU. 
We managed to further increase the performance by $\sim7.5\%$ and reduce the memory consumption by $36.5\%$ by utilizing an AA streaming pattern on GPUs.
We were also able to show a scaling efficiency of the sparse data structure of over 82\% on the JUWELS Booster and LUMI-G HPC system for 1024 and 4096 GPUs, respectively.

We set up a porous media flow simulation and achieved a speed-up of 1.9 and reduced the memory consumption by $50 \%$. 
For an artery blood flow simulation, we gained a speed-up of 2 dependent on the block sizes and achieved a decrease of memory consumption of about 75\%. 
We experienced imbalances in the distribution of the work over the MPI processes when using the sparse data structures. 
To maximize the efficiency of the sparse LBM, we employed load balancing, but more research is needed to fully optimize the workload distribution while maintaining the spatial locality of the neighboring blocks.

To further increase the flexibility of the code generation with \lbmpy{}, future work is planned to support the emerging INTEL GPUs by having a SYCL\endnote{SYCL: \url{https://www.khronos.org/sycl/}, accessed: 2024-04-16} back-end. As SYCL is available on most currently available systems, the code generation could used to generate optimized sparse LB kernels for all existing and upcoming hardware such as CPU, accelerator cards, or even exotic hardware such as Accelerated Processing Units (APUs) or Field Programmable Gate Arrays (FPGAs).

\begin{acks}
This work was supported by the SCALABLE project (\url{www.scalable-hpc.eu/}). This project has received funding from the European High-Performance Computing Joint Undertaking (JU) under grant agreement No 956000. The JU receives support from the European Union’s Horizon 2020 research and innovation program and France, Germany, and the Czech Republic. The authors gratefully acknowledge the Gauss Centre for Supercomputing e.V. (\url{www.gauss-centre.eu}) for funding this project by providing computing time through the John von Neumann Institute for Computing (NIC) on the GCS Supercomputer JUWELS at Jülich Supercomputing Centre (JSC). We acknowledge the EuroHPC Joint Undertaking for awarding this project access to the EuroHPC supercomputer LUMI, hosted by CSC (Finland) and the LUMI consortium through a EuroHPC Regular Access call. The authors gratefully acknowledge the scientific support and HPC resources provided by the Erlangen National High Performance Computing Center (NHR@FAU) of the Friedrich-AlexanderUniversität Erlangen-Nürnberg (FAU).
\end{acks}

\theendnotes

\bibliographystyle{sageh}

\begin{thebibliography}{99}
\providecommand{\natexlab}[1]{#1}
\providecommand{\url}[1]{\texttt{#1}}
\providecommand{\urlprefix}{URL}
\expandafter\ifx\csname urlstyle\endcsname\relax
  \providecommand{\doi}[1]{DOI:\discretionary{}{}{}#1}\else
  \providecommand{\doi}{DOI:\discretionary{}{}{}\begingroup
  \urlstyle{rm}\Url}\fi

\bibitem[{Afrouzi et~al.(2020)Afrouzi, Ahmadian, Hosseini, Arasteh, Toghraie
  and Rostami}]{afrouziSimulationBloodFlow2020a}
Afrouzi HH, Ahmadian M, Hosseini M, Arasteh H, Toghraie D and Rostami S (2020)
  Simulation of blood flow in arteries with aneurysm: {{Lattice Boltzmann
  Approach}} ({{LBM}}).
\newblock \emph{Computer Methods and Programs in Biomedicine} 187: 105312.
\newblock \doi{10.1016/j.cmpb.2019.105312}.

\bibitem[{Alvarez(2021)}]{juwels}
Alvarez D (2021) Juwels cluster and booster: Exascale pathfinder with modular
  supercomputing architecture at juelich supercomputing centre.
\newblock \emph{Journal of large-scale research facilities JLSRF} 7.
\newblock \doi{10.17815/jlsrf-7-183}.

\bibitem[{Ambekar et~al.(2023)Ambekar, Schwarzmeier, R{\"u}de and
  Buwa}]{ambekarParticleResolvedTurbulent2023a}
Ambekar AS, Schwarzmeier C, R{\"u}de U and Buwa VV (2023) Particle-resolved
  turbulent flow in a packed bed: {{RANS}}, {{LES}}, and {{DNS}} simulations.
\newblock \emph{AIChE Journal} 69(1).
\newblock \doi{10.1002/aic.17615}.

\bibitem[{Axner et~al.(2009)Axner, Hoekstra, Jeays, Lawford, Hose and
  Sloot}]{axnerSimulationsTimeHarmonic2009}
Axner L, Hoekstra AG, Jeays A, Lawford P, Hose R and Sloot PM (2009)
  Simulations of time harmonic blood flow in the {{Mesenteric}} artery:
  Comparing finite element and lattice {{Boltzmann}} methods.
\newblock \emph{BioMedical Engineering OnLine} 8(1): 23.
\newblock \doi{10.1186/1475-925X-8-23}.

\bibitem[{Bailey et~al.(2009)Bailey, Myre, Walsh, Lilja and
  Saar}]{baileyAcceleratingLatticeBoltzmann2009}
Bailey P, Myre J, Walsh SD, Lilja DJ and Saar MO (2009) Accelerating {{Lattice
  Boltzmann Fluid Flow Simulations Using Graphics Processors}}.
\newblock In: \emph{2009 {{International Conference}} on {{Parallel
  Processing}}}. pp. 550--557.
\newblock \doi{10.1109/ICPP.2009.38}.

\bibitem[{Bauer et~al.(2020)Bauer, Eibl, Godenschwager, Kohl, Kuron, Rettinger,
  Schornbaum, Schwarzmeier, Thönnes, Köstler and Rüde}]{waLBerla}
Bauer M, Eibl S, Godenschwager C, Kohl N, Kuron M, Rettinger C, Schornbaum F,
  Schwarzmeier C, Thönnes D, Köstler H and Rüde U (2020) walberla: A
  block-structured high-performance framework for multiphysics simulations.
\newblock \doi{10.1016/j.camwa.2020.01.007}.

\bibitem[{Bauer et~al.(2019)Bauer, H\"{o}tzer, Ernst, Hammer, Seiz, Hierl,
  H\"{o}nig, K\"{o}stler, Wellein, Nestler and R\"{u}de}]{Bauer19}
Bauer M, H\"{o}tzer J, Ernst D, Hammer J, Seiz M, Hierl H, H\"{o}nig J,
  K\"{o}stler H, Wellein G, Nestler B and R\"{u}de U (2019) Code generation for
  massively parallel phase-field simulations.
\newblock Association for Computing Machinery.
\newblock \doi{10.1145/3295500.3356186}.

\bibitem[{Bauer et~al.(2021)Bauer, Köstler and Rüde}]{lbmpy}
Bauer M, Köstler H and Rüde U (2021) lbmpy: Automatic code generation for
  efficient parallel lattice {Boltzmann} methods.
\newblock \emph{Journal of Computational Science} 49: 101269.
\newblock \doi{10.1016/j.jocs.2020.101269}.

\bibitem[{Bernaschi et~al.(2010)Bernaschi, Fatica, Melchionna, Succi and
  Kaxiras}]{bernaschiFlexibleHighPerformance2010}
Bernaschi M, Fatica M, Melchionna S, Succi S and Kaxiras E (2010) A flexible
  high-performance {{Lattice Boltzmann GPU}} code for the simulations of fluid
  flows in complex geometries.
\newblock \emph{Concurrency and Computation: Practice and Experience} 22(1):
  1--14.
\newblock \doi{10.1002/cpe.1466}.

\bibitem[{Bernaschi et~al.(2008)Bernaschi, Succi, Fyta, Kaxiras, Melchionna and
  Sircar}]{bernaschiMUPHYParallelHigh2008}
Bernaschi M, Succi S, Fyta M, Kaxiras E, Melchionna S and Sircar JK (2008)
  {{MUPHY}}: {{A}} parallel high performance {{MUlti PHYsics}}/{{Scale}} code.
\newblock In: \emph{2008 {{IEEE International Symposium}} on {{Parallel}} and
  {{Distributed Processing}}}. pp. 1--8.
\newblock \doi{10.1109/IPDPS.2008.4536464}.

\bibitem[{Brodtkorb et~al.(2013)Brodtkorb, Hagen and
  S{\ae}tra}]{brodtkorbGraphicsProcessingUnit2013}
Brodtkorb AR, Hagen TR and S{\ae}tra ML (2013) Graphics processing unit
  ({{GPU}}) programming strategies and trends in {{GPU}} computing.
\newblock \emph{Journal of Parallel and Distributed Computing} 73(1): 4--13.
\newblock \doi{10.1016/j.jpdc.2012.04.003}.

\bibitem[{Chen and Doolen(1998)}]{chenLATTICEBOLTZMANNMETHOD1998a}
Chen S and Doolen GD (1998) {{Lattice Boltzmann method for fluid flows}}.
\newblock \emph{Annual Review of Fluid Mechanics} 30(1): 329--364.
\newblock \doi{10.1146/annurev.fluid.30.1.329}.

\bibitem[{Fattahi et~al.(2016)Fattahi, Waluga, Wohlmuth and
  R{\"u}de}]{fattahiLargeScaleLattice2016}
Fattahi E, Waluga C, Wohlmuth B and R{\"u}de U (2016) Large {{Scale Lattice
  Boltzmann Simulation}} for the {{Coupling}} of {{Free}} and {{Porous Media
  Flow}}.
\newblock In: Kozubek T, Blaheta R, {\v S}{\'i}stek J, Rozlo{\v z}n{\'i}k M and
  {\v C}erm{\'a}k M (eds.) \emph{High {{Performance Computing}} in {{Science}}
  and {{Engineering}}}, volume 9611. Cham: Springer International Publishing.
\newblock ISBN 978-3-319-40360-1 978-3-319-40361-8, pp. 1--18.
\newblock \doi{10.1007/978-3-319-40361-8_1}.

\bibitem[{Geier et~al.(2015)Geier, Sch{\"o}nherr, Pasquali and
  Krafczyk}]{geierCumulantLatticeBoltzmann2015}
Geier M, Sch{\"o}nherr M, Pasquali A and Krafczyk M (2015) The cumulant lattice
  {{Boltzmann}} equation in three dimensions: {{Theory}} and validation.
\newblock \emph{Computers \& Mathematics with Applications} 70(4): 507--547.
\newblock \doi{10.1016/j.camwa.2015.05.001}.

\bibitem[{Godenschwager et~al.(2013)Godenschwager, Schornbaum, Bauer,
  K{\"o}stler and R{\"u}de}]{godenschwagerFrameworkHybridParallel2013}
Godenschwager C, Schornbaum F, Bauer M, K{\"o}stler H and R{\"u}de U (2013) A
  framework for hybrid parallel flow simulations with a trillion cells in
  complex geometries.
\newblock In: \emph{Proceedings of the {{International Conference}} on {{High
  Performance Computing}}, {{Networking}}, {{Storage}} and {{Analysis}}},
  {{SC}} '13. {New York, NY, USA}: {Association for Computing Machinery}.
\newblock ISBN 978-1-4503-2378-9, pp. 1--12.
\newblock \doi{10.1145/2503210.2503273}.

\bibitem[{Han and Cundall(2013)}]{hanLBMModelingFluid2013}
Han Y and Cundall PA (2013) {{LBM}}--{{DEM}} modeling of fluid--solid
  interaction in porous media.
\newblock \emph{International Journal for Numerical and Analytical Methods in
  Geomechanics} 37(10): 1391--1407.
\newblock \doi{10.1002/nag.2096}.

\bibitem[{Hasert et~al.(2014)Hasert, Masilamani, Zimny, Klimach, Qi, Bernsdorf
  and Roller}]{hasertComplexFluidSimulations2014}
Hasert M, Masilamani K, Zimny S, Klimach H, Qi J, Bernsdorf J and Roller S
  (2014) Complex fluid simulations with the parallel tree-based {{Lattice
  Boltzmann}} solver {{Musubi}}.
\newblock \emph{Journal of Computational Science} 5(5): 784--794.
\newblock \doi{10.1016/j.jocs.2013.11.001}.

\bibitem[{Hennig et~al.(2022)Hennig, Holzer and
  R{\"u}de}]{hennigAdvancedAutomaticCode2022}
Hennig F, Holzer M and R{\"u}de U (2022) Advanced {{Automatic Code Generation}}
  for {{Multiple Relaxation-Time Lattice Boltzmann Methods}}.
\newblock \doi{10.48550/arXiv.2211.02435}.

\bibitem[{Hijma et~al.(2023)Hijma, Heldens, Sclocco, {van Werkhoven} and
  Bal}]{hijmaOptimizationTechniquesGPU2023}
Hijma P, Heldens S, Sclocco A, {van Werkhoven} B and Bal HE (2023) Optimization
  {{Techniques}} for {{GPU Programming}}.
\newblock \emph{ACM Computing Surveys} 55(11): 239:1--239:81.
\newblock \doi{10.1145/3570638}.

\bibitem[{Holzer et~al.(2024)Holzer, Mitchell, Leonardi and
  Ruede}]{holzerDevelopmentCentralmomentPhasefield2024}
Holzer M, Mitchell T, Leonardi CR and Ruede U (2024) Development of a
  central-moment phase-field lattice {{Boltzmann}} model for thermocapillary
  flows: {{Droplet}} capture and computational performance.

\bibitem[{Kemmler et~al.(2023)Kemmler, Schwarzmeier, Rettinger, Plewinski,
  Núñez-González, Köstler, Rüde and Vowinckel}]{antidunes}
Kemmler S, Schwarzmeier C, Rettinger C, Plewinski J, Núñez-González F,
  Köstler H, Rüde U and Vowinckel B (2023) {Geometrically} {Resolved}
  {Simulation} of {Upstream} {Migrating} {Antidune} {Formation} and
  {Propagation}.
\newblock In: \emph{40th IAHR World Congres}.

\bibitem[{Kr\"{u}ger et~al.(2017)Kr\"{u}ger, Kusumaatmaja, Kuzmin, Shardt,
  Silva and Viggen}]{lbm_book}
Kr\"{u}ger T, Kusumaatmaja H, Kuzmin A, Shardt O, Silva G and Viggen EM (2017)
  \emph{The Lattice {Boltzmann} Method}.
\newblock Springer International Publishing.
\newblock \doi{10.1007/978-3-319-44649-3}.

\bibitem[{Kummerl{\"a}nder et~al.(2023)Kummerl{\"a}nder, Dorn, Frank and
  Krause}]{kummerlanderImplicitPropagationDirectly2023}
Kummerl{\"a}nder A, Dorn M, Frank M and Krause MJ (2023) Implicit propagation
  of directly addressed grids in lattice {{Boltzmann}} methods.
\newblock \emph{Concurrency and Computation: Practice and Experience}
  \doi{10.1002/cpe.7509}.

\bibitem[{Lai et~al.(2020)Lai, Yu, Tian and Li}]{laiHybridMPICUDA2020}
Lai J, Yu H, Tian Z and Li H (2020) Hybrid {{MPI}} and {{CUDA Parallelization}}
  for {{CFD Applications}} on {{Multi-GPU HPC Clusters}}.
\newblock \emph{Scientific Programming} 2020: e8862123.
\newblock \doi{10.1155/2020/8862123}.

\bibitem[{Latt et~al.(2021)Latt, Malaspinas, Kontaxakis, Parmigiani, Lagrava,
  Brogi, Belgacem, Thorimbert, Leclaire, Li, Marson, Lemus, Kotsalos, Conradin,
  Coreixas, Petkantchin, Raynaud, Beny and
  Chopard}]{lattPalabosParallelLattice2021}
Latt J, Malaspinas O, Kontaxakis D, Parmigiani A, Lagrava D, Brogi F, Belgacem
  MB, Thorimbert Y, Leclaire S, Li S, Marson F, Lemus J, Kotsalos C, Conradin
  R, Coreixas C, Petkantchin R, Raynaud F, Beny J and Chopard B (2021) Palabos:
  {{Parallel Lattice Boltzmann Solver}}.
\newblock \emph{Computers \& Mathematics with Applications} 81: 334--350.
\newblock \doi{10.1016/j.camwa.2020.03.022}.

\bibitem[{Lehmann(2022)}]{lehmannEsotericPullEsoteric2022}
Lehmann M (2022) Esoteric {{Pull}} and {{Esoteric Push}}: {{Two Simple In-Place
  Streaming Schemes}} for the {{Lattice Boltzmann Method}} on {{GPUs}}.
\newblock \emph{Computation} 10(6): 92.
\newblock \doi{10.3390/computation10060092}.

\bibitem[{Lehmann et~al.(2022)Lehmann, Krause, Amati, Sega, Harting and
  Gekle}]{lehmannAccuracyPerformanceLattice2022}
Lehmann M, Krause MJ, Amati G, Sega M, Harting J and Gekle S (2022) On the
  accuracy and performance of the lattice {{Boltzmann}} method with 64-bit,
  32-bit and novel 16-bit number formats.
\newblock \emph{Physical Review E} 106(1): 015308.
\newblock \doi{10.1103/PhysRevE.106.015308}.

\bibitem[{Liu et~al.(2023)Liu, Chu, Lv, Liu, Fu and
  Yang}]{liuAcceleratingLargeScaleCFD2023}
Liu Z, Chu X, Lv X, Liu H, Fu H and Yang G (2023) Accelerating {{Large-Scale
  CFD Simulations}} with {{Lattice Boltzmann Method}} on a 40-{{Million-Core
  Sunway Supercomputer}}.
\newblock In: \emph{Proceedings of the 52nd {{International Conference}} on
  {{Parallel Processing}}}. Salt Lake City UT USA: ACM.
\newblock ISBN 9798400708435, pp. 797--806.
\newblock \doi{10.1145/3605573.3605605}.

\bibitem[{Martin et~al.(2023)Martin, Liu, Ladd, Lee, Gounley, Vetter, Patel,
  Rizzi, Mateevitsi, Insley and
  Randles}]{martinPerformanceEvaluationHeterogeneous2023}
Martin A, Liu G, Ladd W, Lee S, Gounley J, Vetter J, Patel S, Rizzi S,
  Mateevitsi V, Insley J and Randles A (2023) Performance {{Evaluation}} of
  {{Heterogeneous GPU Programming Frameworks}} for {{Hemodynamic Simulations}}.
\newblock In: \emph{Proceedings of the {{SC}} '23 {{Workshops}} of {{The
  International Conference}} on {{High Performance Computing}}, {{Network}},
  {{Storage}}, and {{Analysis}}}. Denver CO USA: ACM.
\newblock ISBN 9798400707858, pp. 1126--1137.
\newblock \doi{10.1145/3624062.3624188}.

\bibitem[{Mattila et~al.(2016)Mattila, Puurtinen, Hyv{\"a}luoma, Surmas,
  Myllys, Turpeinen, Roberts{\'e}n, Westerholm and
  Timonen}]{mattilaProspectComputingPorous2016a}
Mattila K, Puurtinen T, Hyv{\"a}luoma J, Surmas R, Myllys M, Turpeinen T,
  Roberts{\'e}n F, Westerholm J and Timonen J (2016) A prospect for computing
  in porous materials research: {{Very}} large fluid flow simulations.
\newblock \emph{Journal of Computational Science} 12: 62--76.
\newblock \doi{10.1016/j.jocs.2015.11.013}.

\bibitem[{Pan et~al.(2004)Pan, Prins and
  Miller}]{panHighperformanceLatticeBoltzmann2004}
Pan C, Prins JF and Miller CT (2004) A high-performance lattice {{Boltzmann}}
  implementation to model flow in porous media.
\newblock \emph{Computer Physics Communications} 158(2): 89--105.
\newblock \doi{10.1016/j.cpc.2003.12.003}.

\bibitem[{Pearson(2023)}]{pearsonInterconnectBandwidthHeterogeneity2023}
Pearson C (2023) Interconnect {{Bandwidth Heterogeneity}} on {{AMD MI250x}} and
  {{Infinity Fabric}}.

\bibitem[{Rak(2024)}]{rakParallelProgrammingHybrid2024}
Rak T (2024) Parallel {{Programming}} in the {{Hybrid Model}} on the {{HPC
  Clusters}}.
\newblock In: Malhotra R, Sumalatha L, Yassin SMW, Patgiri R and Muppalaneni NB
  (eds.) \emph{High {{Performance Computing}}, {{Smart Devices}} and
  {{Networks}}}. Singapore: Springer Nature.
\newblock ISBN 978-981-9966-90-5, pp. 207--219.
\newblock \doi{10.1007/978-981-99-6690-5_15}.

\bibitem[{Randles et~al.(2015)Randles, Draeger, Oppelstrup, Krauss and
  Gunnels}]{randlesMassivelyParallelModels2015}
Randles A, Draeger EW, Oppelstrup T, Krauss L and Gunnels JA (2015) Massively
  parallel models of the human circulatory system.
\newblock In: \emph{Proceedings of the {{International Conference}} for {{High
  Performance Computing}}, {{Networking}}, {{Storage}} and {{Analysis}}},
  {{SC}} '15. New York, NY, USA: Association for Computing Machinery.
\newblock ISBN 978-1-4503-3723-6, pp. 1--11.
\newblock \doi{10.1145/2807591.2807676}.

\bibitem[{Randles et~al.(2013)Randles, Kale, Hammond, Gropp and
  Kaxiras}]{randlesPerformanceAnalysisLattice2013a}
Randles AP, Kale V, Hammond J, Gropp W and Kaxiras E (2013) Performance
  {{Analysis}} of the {{Lattice Boltzmann Model Beyond Navier-Stokes}}.
\newblock In: \emph{2013 {{IEEE}} 27th {{International Symposium}} on
  {{Parallel}} and {{Distributed Processing}}}. Cambridge, MA, USA: IEEE.
\newblock ISBN 978-1-4673-6066-1 978-0-7695-4971-2, pp. 1063--1074.
\newblock \doi{10.1109/IPDPS.2013.109}.

\bibitem[{Rettinger and R{\"u}de(2018)}]{rettingerCoupledLatticeBoltzmann2018}
Rettinger C and R{\"u}de U (2018) A coupled lattice {{Boltzmann}} method and
  discrete element method for discrete particle simulations of particulate
  flows.
\newblock \emph{Computers \& Fluids} 172: 706--719.
\newblock \doi{10.1016/j.compfluid.2018.01.023}.

\bibitem[{Schornbaum and
  R{\"u}de(2016)}]{schornbaumMassivelyParallelAlgorithms2016}
Schornbaum F and R{\"u}de U (2016) Massively {{Parallel Algorithms}} for the
  {{Lattice Boltzmann Method}} on {{Non-uniform Grids}}.
\newblock \emph{SIAM Journal on Scientific Computing} 38(2): C96--C126.
\newblock \doi{10.1137/15M1035240}.

\bibitem[{Schornbaum and
  R{\"u}de(2018)}]{schornbaumExtremeScaleBlockStructuredAdaptive2018}
Schornbaum F and R{\"u}de U (2018) Extreme-{{Scale Block-Structured Adaptive
  Mesh Refinement}}.
\newblock \emph{SIAM Journal on Scientific Computing} 40(3): C358--C387.
\newblock \doi{10.1137/17M1128411}.

\bibitem[{Schulz et~al.(2002)Schulz, Krafczyk, T{\"o}lke and
  Rank}]{schulzParallelizationStrategiesEfficiency2002}
Schulz M, Krafczyk M, T{\"o}lke J and Rank E (2002) Parallelization
  {{Strategies}} and {{Efficiency}} of {{CFD Computations}} in {{Complex
  Geometries Using Lattice Boltzmann Methods}} on {{High-Performance
  Computers}}.
\newblock In: Griebel M, Keyes DE, Nieminen RM, Roose D, Schlick T, Breuer M,
  Durst F and Zenger C (eds.) \emph{High {{Performance Scientific And
  Engineering Computing}}}, volume~21. Berlin, Heidelberg: Springer Berlin
  Heidelberg.
\newblock ISBN 978-3-540-42946-3 978-3-642-55919-8, pp. 115--122.
\newblock \doi{10.1007/978-3-642-55919-8_13}.

\bibitem[{Siefert et~al.(2023)Siefert, Pearson, Olivier, Prokopenko, Hu and
  Fuller}]{siefertLatencyBandwidthMicrobenchmarks2023a}
Siefert CM, Pearson C, Olivier SL, Prokopenko A, Hu J and Fuller TJ (2023)
  Latency and {{Bandwidth Microbenchmarks}} of {{US Department}} of {{Energy
  Systems}} in the {{June}} 2023 {{Top}} 500 {{List}}.
\newblock In: \emph{Proceedings of the {{SC}} '23 {{Workshops}} of {{The
  International Conference}} on {{High Performance Computing}}, {{Network}},
  {{Storage}}, and {{Analysis}}}. Denver CO USA: ACM.
\newblock ISBN 9798400707858, pp. 1298--1305.
\newblock \doi{10.1145/3624062.3624203}.

\bibitem[{Spinelli et~al.(2023)Spinelli, Horstmann, Masilamani, Soni, Klimach,
  St{\"u}ck and Roller}]{spinelliHPCPerformanceStudy2023}
Spinelli GG, Horstmann T, Masilamani K, Soni MM, Klimach H, St{\"u}ck A and
  Roller S (2023) {{HPC}} performance study of different collision models using
  the {{Lattice Boltzmann}} solver {{Musubi}}.
\newblock \emph{Computers \& Fluids} 255: 105833.
\newblock \doi{10.1016/j.compfluid.2023.105833}.

\bibitem[{Vidal et~al.(2010)Vidal, Roy and
  Bertrand}]{vidalImprovingPerformanceLarge2010}
Vidal D, Roy R and Bertrand F (2010) On improving the performance of large
  parallel lattice {{Boltzmann}} flow simulations in heterogeneous porous
  media.
\newblock \emph{Computers \& Fluids} 39(2): 324--337.
\newblock \doi{10.1016/j.compfluid.2009.09.011}.

\bibitem[{Wang et~al.(2005)Wang, Zhang, Bengough and
  Crawford}]{wangDomaindecompositionMethodParallel2005}
Wang J, Zhang X, Bengough AG and Crawford JW (2005) Domain-decomposition method
  for parallel lattice {{Boltzmann}} simulation of incompressible flow in
  porous media.
\newblock \emph{Physical Review E} 72(1): 016706.
\newblock \doi{10.1103/PhysRevE.72.016706}.

\bibitem[{Watanabe and Hu(2022)}]{watanabePerformanceEvaluationLattice2022}
Watanabe S and Hu C (2022) Performance {{Evaluation}} of {{Lattice Boltzmann
  Method}} for {{Fluid Simulation}} on {{A64FX Processor}} and {{Supercomputer
  Fugaku}}.
\newblock In: \emph{International {{Conference}} on {{High Performance
  Computing}} in {{Asia-Pacific Region}}}. Virtual Event Japan: ACM.
\newblock ISBN 978-1-4503-8498-8, pp. 1--9.
\newblock \doi{10.1145/3492805.3492811}.

\bibitem[{Wittmann et~al.(2013)Wittmann, Zeiser, Hager and
  Wellein}]{wittmannComparisonDifferentPropagation2013}
Wittmann M, Zeiser T, Hager G and Wellein G (2013) Comparison of different
  propagation steps for lattice {{Boltzmann}} methods.
\newblock \emph{Computers \& Mathematics with Applications} 65(6): 924--935.
\newblock \doi{10.1016/j.camwa.2012.05.002}.

\bibitem[{Yang et~al.(2023)Yang, Chen, Chen and
  Wang}]{yangImplementationDirectaddressingBased2023}
Yang G, Chen Y, Chen S and Wang M (2023) Implementation of a direct-addressing
  based lattice {{Boltzmann GPU}} solver for multiphase flow in porous media.
\newblock \emph{Computer Physics Communications} 291: 108828.
\newblock \doi{10.1016/j.cpc.2023.108828}.

\bibitem[{Zeiser et~al.(2009)Zeiser, Hager and
  Wellein}]{zeiserBENCHMARKANALYSISAPPLICATION2009}
Zeiser T, Hager G and Wellein G (2009) {{Benchmark analysis and application
  results for lattice Boltzmann simulations on NEC SX vector and Intel Nehalem
  systems.}}
\newblock \emph{Parallel Processing Letters} 19(04): 491--511.
\newblock \doi{10.1142/S0129626409000389}.

\end{thebibliography}
\end{document}